\documentclass[a4paper,11pt]{article}

\usepackage{jheppub}
\usepackage{physics}
\usepackage{braket}
\usepackage{tikz}
\usepackage{todonotes}
\usetikzlibrary{decorations.pathreplacing}
\usepackage{faktor}
\usepackage{enumerate}
\usepackage{slashed}

\newcommand{\In}{\text{in}}
\newcommand{\Out}{\text{out}}

\newcommand{\eff}{\text{eff}}
\DeclareMathOperator{\Span}{Span}
\DeclareMathOperator{\Vol}{Vol}

\title{Self-normalizing Path Integrals}

\author[a]{Ivan M. Burbano}
\author[b]{and Francisco Calderón}

\affiliation[a]{Department of Physics, University of California, Berkeley\\
366 Physics North MC
7300, Berkeley, CA 94720-7300, USA}

\affiliation[b]{Department of Philosophy, University of Michigan\\
435 South State Street
2215 Angell Hall, Ann Arbor, MI 48109-1003, USA}

\emailAdd{ivan\_burbano@berkeley.edu}
\emailAdd{fcalder@umich.edu}

\abstract{The normalization in the path integral approach to quantum field theory, in contrast with statistical field theory, can contain physical information. The main claim of this paper is that the inner product on the space of field configurations, one of the fundamental pieces of data required to be added to quantize a classical field theory, determines the normalization of the path integral. In fact, dimensional analysis shows that the introduction of this structure necessarily introduces a scale that is left unfixed by the classical theory. We study the dependence of the theory on this scale. This allows us to explore mechanisms that can be used to fix the normalization based on cutting and gluing different integrals. ``Self-normalizing'' path integrals, those independent of the scale, play an important role in this process. Furthermore, we show that the scale dependence encodes other important physical data: we use it to give a conceptually clear derivation of the chiral anomaly. Several explicit examples, including the scalar and compact bosons in different geometries, supplement our discussion.}

\dedicated{Dedicated to Father Francis Wehri}

\begin{document}

\maketitle

\section{Introduction}\label{sec:intro}

Path integral formulations of quantum field theory (QFT) allow us to answer how a quantum superposition of states evolves in time \cite{Feynman2005a}. By assigning phases to each possible field trajectory, we have a ``quantization'' of the classical version of the theory. A weighted sum of these phases, obtained from the theory's classical action, provides the time evolution of a quantum state. Similar sums appear in statistical field theory (SFT), where the probability of each field configuration is proportional to a Boltzmann weight, consisting of an exponential suppressed by the energy of the configuration. Sums of these Boltzmann weights are then used to compute expectation values for different observables.

The appearance of path integrals in both fields suggests a relationship between them. Once we identify spacetime and space,\footnote{We are focusing on equilibrium SFT, where the probability distribution of the field configurations has no time dependence.} trajectories and configurations, actions and energies, in QFT and SFT, respectively, one arrives at the same mathematical problem. This is, however, not an exact duality \cite{fraser_development_2020}: while the SFT is insensitive to a rescaling of the partition function, the normalization on the QFT side will be mediated by a probabilistic interpretation of the time evolution of the weights we assign to all paths, whose sum cannot be greater than one. In this paper, we will pinpoint the origin of this normalization within the foundations of the path integral formulation. We will also discuss some methods that can be used to explore it and, in certain cases, fix it. 

In QFT, the path integral is a method of \emph{quantization}. Starting from a classical field theory, it provides some general guidelines that can be used to attempt the construction of a quantum theory. These guidelines are, however, not a mapping, and the program of assigning to each classical theory a unique quantum mechanical one is bound to fail \cite{Groenewold1946}. Instead, the construction of a quantum theory requires additional data. Amongst these, one must choose an inner product on the space of field trajectories \cite{Polchinski1986}. The main claim of this paper is that this inner product determines the normalization of the path integral.

As we will see through dimensional analysis, this inner product requires introducing a scale $\mu$. This is analogous to the auxiliary energy scale $\mu$ that is usually introduced in dimensional regularization \cite{Collins1984}, or to the auxiliary phase space area scale $h$ used in the classical treatment of the statistics of the ideal gas. In this paper, we will use zeta function regularization, supplemented by heat-kernel methods \cite{Ramond1990}, to flesh out the dependence of path integrals with respect to this scale. We will do this by introducing a \emph{beta function} determining the dependence on this scale of the associated effective action.

One of the major tools we will use in the determination of the $\mu$-dependence stems from a very simple physical idea: given three instants in time $t_1<t_2<t_3$, understanding the time evolution on the intervals $[t_1,t_2]$ and $[t_2,t_3]$ fully determines the time evolution in the whole interval $[t_1,t_3]$. More generally, a path integral in a given spacetime can be understood by cutting the spacetime into patches. Given that the beta function depends on the geometry on which the path integral is computed, each patch will have a different dependence on $\mu$. In fact, a comparison of the $\mu$ dependence of the path integrals in these different geometries should allow one to fix the value of $\mu$ (see \cite{Bentov2021} for an application of this in a free, $D=1$ theory). 

We will further highlight the possibility that some integrals might, in fact, be independent of $\mu$ (see, e. g., the path integrals appearing in \cite{Hori2003}). We call these \textit{self-normalizing}. Generically, this happens through a cancellation of the $\mu$-dependence coming from the integrals over quantum fluctuations and the one coming from the instantons of the theory. These are of particular interest to the $\mu$-fixing program described above, for the existence of a self-normalizing parent integral can be used to fix $\mu$ in the non-self-normalizing patches.

This paper is organized as follows. In section \ref{sec:gaussian}, we will review the standard lore behind the computation of Gaussian path integrals on vector spaces of fields \cite{Hawking1977}. (Hawking had already identified the appearance of $\mu$ through different means). We will introduce zeta-function regularization and show the necessity of the scale $\mu$ here as well. In section \ref{sec:bundles}, we extend the framework of section \ref{sec:gaussian} to the case where the space of fields has a non-trivial geometry but can still be expressed in terms of vector bundles. This generalization is necessary to study theories with non-trivial instantons. In section \ref{sec:beta}, we introduce the $\mu$ beta function and the heat-kernel techniques we will use to compute it. In section \ref{sec:interpretation}, we study the behavior of path integrals under the cutting and pasting of spacetimes, guided by probabilistic quantum principles. We argue that a complete understanding of these operations should allow one to fix the value of $\mu$. In section \ref{sec:leveraging}, we give further heuristics indicating that such an understanding might allow one to fix $\beta$ without actually computing any bulk path integrals. We give explicit evidence of this in the $D=1$ case through an ``index-like'' theorem. Finally, in section \ref{sec:examples}, we give several examples, in varying degrees of generality, that exhibit the techniques developed in the body of the paper.

For the examples, we first discuss the free scalar field on general closed manifolds. In particular, we show that in odd dimensions (such as in the computation of the partition function of a quantum harmonic oscillator), the path integral self-normalizes. For even dimensions, we provide a general formula for the beta function in terms of the geometry of the manifold. We also give explicit formulae in terms of lattice series for the full path integral on $D$-dimensional tori. We then turn to the study of manifolds with boundary, focusing on cylinders, on which we give a general formula for the beta function as well. On cylinders with toric lids, we further provide a method for determining the path integral in terms of toric path integrals. Throughout, we give explicit formulae for the $D=1$ case, where we are able to recover the well-known results for the harmonic oscillator. In this case, we are able to explicitly show how to fix $\mu$ through the methods described in the body of the paper. We further provide an argument, first suggested to us by Bruno de S. L. Torres \cite{bruno2023}, indicating that the value of $\mu$ found in this case may also be the correct value in higher dimensions.

We also consider the case of a compact boson (the $O(2)$ non-linear sigma model). We show that in odd-dimensional tori, the path integral self-normalizes as well. We further provide explicit formulae for the partition functions and transition amplitudes in the $D=1$ case and show that $\mu$ acquires the same value as on the free scalar field.

Finally, we consider the chiral anomaly present in the quantization of a Dirac fermion in a background electromagnetic field. From our point of view, the anomaly arises due to chiral transformations not preserving the fermionic inner product (for it looks like a mass term). This is quantified by a non-trivial change of the scale $\mu$. We can compute this as a change in the beta function of the theory. In particular, our methods provide a simple proof that the one-loop result is, in fact, exact.

Before we start, we recommend to the reader that the main body of the paper be read in conjunction with the examples. Having the $D=1$ case (which corresponds to standard particle mechanics) in mind at all times can be particularly useful if one wishes to understand the main physical points without going through difficult computations.

\section{Gaussian Path Integrals}\label{sec:gaussian}

We will now give a self-contained review of the main computation scheme described in \citep{Hawking1977}. Let $\mathcal{E}$ be a vector space of fields denoted by $\phi$ in a $D$-dimensional Euclidean spacetime $M$. Let us assume we have a quadratic action $S(\phi)$ corresponding to a free theory. Our goal is then to define the path integral
\begin{equation}\label{eq:path_integral}
    \int_\mathcal{E}\mathcal{D}\phi\, e^{-S(\phi)}.
\end{equation}
In order to do this, we will attempt to extend the finite-dimensional result 
\begin{equation}\label{eq:gaussian_finite}
    \int_{V}\dd[N]{\vec{x}} e^{-\frac{1}{2}\ev{\vec{x}, 2\pi A\vec{x}}}=\det(A)^{-1/2},
\end{equation}
valid for finite dimensional vector spaces $V$, equipped with an inner product $\ev{\cdot,\cdot}$, its associated volume form $\dd[N]{\vec{x}}$, and a positive symmetric operator $A$. This program necessarily involves formal manipulations of ill-defined quantities. One hopes that these will pave the way to a rigorous definition of \eqref{eq:path_integral}, which can then be used to derive further properties of the theory. But before doing this, let us guide our search by providing some useful comments to keep in mind when doing this:

\begin{enumerate}
    \item In the classical theory, the space $\mathcal{E}$ is not equipped with an inner product. This inner product is, however, fundamental in \eqref{eq:gaussian_finite}. In fact, in the finite-dimensional case, it defines the measure $\dd[N]{x}$. Thus, it would seem reasonable that such an inner product on $\mathcal{E}$ would be instrumental in defining $\mathcal{D}\phi$. Furthermore, while the mathematical standpoint of such a measure remains obscure at this stage, it is clear what an inner product on $\mathcal{E}$ would mean. We will thus posit that such an inner product is part of the data required to quantize the classical theory.
    \item The inner product will also play another role in the path integral quantization. It allows us to express the quadratic form $S$ of the classical theory in terms of an operator $A$ on $\mathcal{E}$ for which $S(\phi)=\ev{\phi, 2\pi A\phi}$.
    \item Having defined the infinite-dimensional operator $A$, one inevitably faces the task of computing infinite-dimensional determinants to employ \eqref{eq:gaussian_finite}. The theory of such determinants is well-developed and employs zeta function regularization. While we will review this below, let us make a remark that will play a fundamental role throughout our discussion. The definition of $\det A$ will involve understanding $A^{-s}$ for $s\in\mathbb{C}$. This, in turn, is generically understood through $\ln A$. For this object to be physically meaningful, we must have a dimensionless operator $A$. Since the action is dimensionless, inner products $\ev{\phi_1,\phi_2}$ must be dimensionless. The definition of such an inner product thus requires introducing a scale $\mu$. For example, the energy of a scalar field in $D$ spacetime dimensions is $[\phi]=(D-2)/2$. Thus, we have a dimensionless inner product given by
    \begin{equation}
        \ev{\phi_1,\phi_2}=\frac{1}{2\pi\mu}\int\dd[D]{x}\sqrt{g}\phi_1\phi_2,
    \end{equation}
    if $[\mu]=-2$. We will thus, in general, not obtain a single result for \eqref{eq:path_integral}, but rather a family of results depending on the scale $\mu$.
\end{enumerate}

Accordingly, we will equip our infinite-dimensional space of fields $\mathcal{E}$ with an inner product $\ev{\cdot,\cdot}$. We then assume that we can find an operator $Q:\mathcal{E}\rightarrow\mathcal{E}$, which is symmetric with respect to our bilinear form,
and for which
\begin{equation}\label{eq:action_operator}
    S(\phi)=\frac{1}{2}\ev{\phi,2\pi \mu Q\phi}.
\end{equation}
Here, $\mu$ is a scale with dimensions $[\mu]=-[Q]$ (i. e., $\sqrt{\mu}$  is a length scale for a bosonic theory, while $\mu$ is a length scale for a fermionic theory). Such a scale can be introduced by taking advantage of the arbitrariness of the normalization of the bilinear form
\begin{equation}
    \ev{\cdot,\cdot}\propto \frac{1}{2\pi\mu}\int\dd[D]{x},
\end{equation}
and guarantees that the operator $A=\mu Q$ is dimensionless. Our first attempt at a definition of \eqref{eq:path_integral} would then be
\begin{equation}\label{eq:gaussian_infinite}
    \int_\mathcal{E}\mathcal{D}\phi\, e^{-S(\phi)}=\det(\mu Q)^{-1/2}.
\end{equation}

The problem has now been reduced to the computation of the latter determinant. It is based on the formal manipulations
\begin{equation}\label{eq:zeta_function_operator}
    \det A=e^{\tr\ln A}=e^{-\tr\left.\dv{s}A^{-s}\right|_{s=0}}=e^{-\zeta'_A(0)}\qc\zeta_A(s):=\tr A^{-s},
\end{equation}
using the relationship between the determinant and the trace, as well as the linearity of the latter, both of which would be completely rigorous for positive operators in the finite-dimensional setting. Fortunately, the mathematical theory of infinite-dimensional determinants is well developed \cite{elizaldeZetaRegularizationTechniques1994, elizaldeTenPhysicalApplications2012}, with the definition given by the last equality in \eqref{eq:zeta_function_operator}. The function $\zeta_A$ defined in \eqref{eq:zeta_function_operator} is known as the zeta function of $A$. This definition only converges for $\Re s$ sufficiently big, but one can analytically extend it in a unique way to $s=0$ with the help of the Mellin transform
\begin{equation}\label{eq:mellin_transform}
    A^{-s}=\frac{1}{\Gamma(s)}\int_0^\infty\dd{t}t^{s-1}e^{-tA},
\end{equation}
which can be checked through the spectral representation of $A$. Along with \eqref{eq:gaussian_infinite}, this defines our initial path integral in terms of an effective action that generically depends on the scale $\mu$.
\begin{equation}\label{eq:gaussian_definition}
    \int_\mathcal{E}\mathcal{D}\phi\, e^{-\frac{1}{\hbar}S(\phi)}=e^{-S_\text{eff}(\mu)}\qc S_\text{eff}(\mu)=\frac{1}{2}\ln\det(\mu Q)=-\frac{1}{2}\zeta'_{\mu Q}(0).
\end{equation}

\section{Path Integrals on Vector Bundles}\label{sec:bundles}

The path integrals interesting for physical applications are often not in the Gaussian form \eqref{eq:path_integral}. On the one hand, one can be interested in the case of interacting theories. On the other, the space of fields $\mathcal{E}$ may fail to be linear. There are at least two reasons for its non-linearity. First, the spacetime of the theory might have boundaries. For instance, in the case of particle mechanics, the space of fields consists of all trajectories that start and end at some fixed events in spacetime, but the sum of two such trajectories will generically not satisfy the appropriate boundary conditions set by these events. In fact, such a sum over spacetime points is meaningless without a distinguished origin, which leads us to the second way $\mathcal{E}$ may fail to be linear: the space in which the field takes values could itself be non-linear. This is the case of particles or strings moving in (non-flat) spacetimes and their associated non-linear sigma models.

However, we can extend most of the formalism developed in the previous section to the case in which $\mathcal{E}$ admits the structure of an infinite-rank vector bundle over a finite-dimensional base space $\mathcal{J}$. Notice that working with vector bundles is a straightforward generalization of fields on vector spaces that preserves the advantages of linearity. We will further demand that this base space is the space of ``instantons'' of the theory (this is similar to the decompositions studied in the BV-BFV formalism of path integrals with boundary \cite{Cattaneo2020}, where they are known as residual fields), that is, the space of solutions to the classical equations of motion associated with $S$. For each instanton $\phi_s\in\mathcal{J}$, its fiber
\begin{equation}
    \mathcal{E}_{\phi_s}=\Set{\phi\in\mathcal{E}|\pi(\phi)=\phi_s},
\end{equation}
will be the space of quantum fluctuations around $\phi_s$. The choice of $\pi:\mathcal{E}\rightarrow\mathcal{J}$ (that is, the choice of quantum fluctuations) is not entirely fixed by the formalism, and the difficulty in the computation of the path integral may greatly vary depending on which $\pi$ is chosen.

We will further embed $\mathcal{E}$ inside a bigger vector space of fields $\mathcal{V}$. An inner product on $\mathcal{V}$ then induces a measure $\mathcal{D}\phi_s$ on $\mathcal{J}$ which one can use to foliate the path integral by its fibers
\begin{equation}
    \int_\mathcal{E}\mathcal{D}\phi\, e^{-S(\phi)}:=\int_\mathcal{J}\mathcal{D}\phi_s\,e^{-S(\phi_s)}\int_{\mathcal{E}_{\phi_s}}\mathcal{D}\tilde{\phi}\,e^{-\int\dd[D]{x}\dd[D]{y}\tilde{\phi}(x)\left.\frac{\delta S}{\delta\phi(x)\delta\phi(y)}\right|_{\phi=\phi_s}\tilde{\phi}(y)}.
\end{equation}
Since $\phi_s$ is in the critical locus of $S$, there is no term linear in the quantum fluctuations $\tilde{\phi}\in\mathcal{E}_{\phi_s}$. The truncation at the second derivative of the action corresponds to the 1-loop, i. e. $\order{\hbar}$ computation of the path integral. Thanks to this foliation and the truncation to this loop order, the fiber-wise integrals are now Gaussian! We can thus introduce an inner product to relate the distribution associated with this second derivative to a linear operator, perform zeta-function regularization on it, and define the effective actions. This leaves us with a finite-dimensional integral on the base space
\begin{equation}
    \int_\mathcal{E}\mathcal{D}\phi\, e^{-S(\phi)}=\int_\mathcal{J}\mathcal{D}\phi_s\,e^{-S_{\text{eff}}(\phi_s,\mu)}\qc S_\text{eff}(\phi_s,\mu)=S(\phi_s)-\frac{1}{2}\zeta_{\mu Q}'(0),
\end{equation}
with $\mu Q$ the operator associated to the second derivative of $S$ by the inner product on $\mathcal{E}$. One expects the latter to generically depend on a scale $\mu$ for the same reasons discussed in the previous section. Furthermore, one should note that, in general, $Q$ will itself depend on $\phi_s$ because this is where the second derivative of the action is being computed. The only exception is when the theory is free and, thus, the second derivative is constant.

\section{Beta Functions and Scale Dependence}\label{sec:beta}

Our definition \eqref{eq:gaussian_definition} contains an ambiguity parametrized by the scale $\mu$. To make this ambiguity explicit, let us compare the result obtained with the energy scale $\mu$ to one obtained with the energy scale $\tilde{\mu}$. Note that $\zeta_{\tilde{\mu} Q}(s)=\qty(\tilde{\mu}/\mu)^{-s}\zeta_{\mu Q}(s)$. This allows us to obtain a beta function for the effective action
\begin{equation}\label{eq:beta}
    \beta:=\left.\dv{S_{\text{eff}}}{\ln\tilde{\mu}}\right|_{\tilde{\mu}=\mu}=\frac{1}{2}\zeta_{\mu Q}(0).
\end{equation}
This, in turn, can be easily computed from \eqref{eq:mellin_transform}. Since $\Gamma(s)\rightarrow\pm\infty$ for $s\rightarrow 0$, we need to understand the divergence in the trace of the integrand of \eqref{eq:mellin_transform} to understand $\zeta_A(0)$. For large $t$, this integral is exponentially suppressed, so the divergence comes from the $t\rightarrow 0$ behavior. Indeed, in this limit, the heat kernel trace diverges as the dimension of $\mathcal{E}$. From an asymptotic expansion of this trace
\begin{equation}\label{eq:asymptotic_heat_kernel}
    \tr(e^{-tA})\sim\sum_{n}a_nt^n,
\end{equation}
we have
\begin{equation}
    \Gamma(s)\zeta_A(s)\sim \sum_na_n\int_0^1\dd{t}t^{s-1+n}+\cdots=\sum_n\frac{a_n}{s+n}+\cdots,
\end{equation}
after focusing on a region of integration where the large-$t$ behavior is unimportant. Therefore, the right-hand side poles are at the values of $n$ in the expansion \eqref{eq:asymptotic_heat_kernel}. Furthermore, since
\begin{equation}
    \lim_{s\rightarrow 0}\frac{1}{\Gamma(s)(s+n)}=\begin{cases}
        1 & n=0,\\
        0 & n\neq 0,
    \end{cases}
\end{equation}
we conclude that $\zeta_A(0)=a_0$ is the constant coefficient of \eqref{eq:asymptotic_heat_kernel}. This entails that the beta function \eqref{eq:beta} is independent of $\mu$ and we can without harm set $\zeta_Q(0):=\zeta_{\mu Q}(0)$. 

In order to fully spell out the dependence of the path integral on $\mu$, we must also consider the dependence of the measure $\mathcal{D}\phi_s$ on this scale. In order to do this, we need to review the way that inner products define integration measures. Consider coordinates $\{c_n\}_{n\in I}$ on $\mathcal{E}$, where
\begin{equation}
    \phi=\sum_{n\in I}c_n\phi_n\qc\ev{\phi_n,\phi_m}_\mu=\delta_{nm}.
\end{equation}
Then, the integration measure can heuristically be written as
\begin{equation}
    (\mathcal{D}\phi)_\mu=\prod_{n\in I}\dd{c_n}.
\end{equation}
To make the dependence on $\mu$ explicit, let us compare this to a theory with a different energy scale $\tilde{\mu}$. We would then have
\begin{equation}
    \ev{\phi_n,\phi_m}_{\mu}=\frac{\tilde{\mu}}{\mu}\ev{\phi_n,\phi_m}_{\tilde{\mu}}
\end{equation}
and the integration measure would be
\begin{equation}
    (\mathcal{D}\phi)_{\tilde{\mu}}=\prod_{n\in I}\qty(\sqrt{\mu/\tilde{\mu}}\dd c_n)=(\mathcal{D}\phi)_\mu e^{-V(\mu)},
\end{equation}
where
\begin{equation}
    V=-\frac{1}{2}\sum_{n\in I}\ln(\mu/\tilde{\mu})
\end{equation}
From this point of view, $\mu$ can be understood as part of an infinite vacuum \citep{Chen2018} counterterm that has to be added to the action to define the quantum theory. In a free theory, these are the only counterterms one would expect since there are no divergent Feynman diagrams.

Restricting our discussion to $\mathcal{J}$, we see that
\begin{equation}\label{eq:measure_beta}
    (\mathcal{D}\phi_s)_{\tilde{\mu}}\propto\qty(\frac{\mu}{\tilde{\mu}})^{\dim\mathcal{J}/2}(\mathcal{D}\phi_s)_\mu,
\end{equation}
for each coordinate associated with an instanton contributes a factor of $\sqrt{\mu/\tilde{\mu}}$. We then conclude that the $\mu$ dependence of the path integral is
\begin{equation}\label{eq:multiplicative_constant}
    \int_\mathcal{E}(\mathcal{D}\phi\,)_{\tilde{\mu}} e^{-S(\phi)}= \qty(\frac{\mu}{\tilde{\mu}})^{\beta+\dim\mathcal{J}/2}\int_\mathcal{E}(\mathcal{D}\phi\,)_{\mu} e^{-S(\phi)}.
\end{equation}

\section{How a Quantum Mechanical Interpretation of the Path Integral Helps Fix \texorpdfstring{$\mu$}{mu}}\label{sec:interpretation}

Having spelled out the dependence on $\mu$ of these path integrals, we should mention the evident elephant in the room: since these path integrals generically depend on this arbitrary scale $\mu$, they cannot correspond to observable physical quantities. Indeed, this is the case in statistical field theory, where these path integrals correspond to the partition function of the theory, which is undetermined up to a multiplicative constant that can absorb factors like the one found in \eqref{eq:multiplicative_constant}. However, in quantum field theory, these path integrals can be directly related to transition amplitudes. Thus, we must clarify how these ambiguous quantities can be connected to such observables.

\begin{figure}[ht]
	\centering
	\begin{tikzpicture}
		\draw plot [smooth, tension = 1] coordinates {(0,0.5)(1,1)(2.3,0.8)(4,1.3)(4,0)(2.5,-0.7)(1,-0.5)(-1,-1)(0,0.5)};
        \draw (-1.04,-0.95) .. controls (-0.8,-1.5) and (-1, -2.5) .. (-0.85,-3.4);
	    \draw (4.33,0.7) .. controls (4,-1.5) and (4, -2.5) .. (4,-3.4);
        \draw plot [smooth, tension = 1] coordinates{(-0.5,-3.2)(1,-2.7)(2,-3)(3,-2.8)(4,-3.4)(3,-3.5)(2,-3.8)(1,-4)(-0.7,-3.6)(-0.5,-3.2)}; 
        \node at (2,0.2) {$\Sigma_\Out$};
        \node at (1,-3.3) {$\Sigma_\In$};
        \node at (1.5,-1.8) {$M$};
    \end{tikzpicture}
	\caption{\label{fig:cobordism}A spacetime $M$ with an incoming manifold $\Sigma_\In$ and an outgoing manifold $\Sigma_\Out$.}
\end{figure}

Suppose we have a spacetime $M$ on which our fields are defined, equipped with a decomposition of its boundary $\partial M$ into an incoming boundary $\Sigma_\text{in}$ and an outgoing boundary $\Sigma_\text{out}$, like the one shown in figure \ref{fig:cobordism}. In many cases, one can understand the Hilbert space $\mathcal{H}(\Sigma)$ on a codimension 1 submanifold $\Sigma\subseteq M$, thought of as a constant time slice in $M$, as the space spanned orthonormally by $\mathcal{E}_\Sigma$, the space of field configurations $\varphi$ on $\Sigma$.\footnote{More precisely $\mathcal{H}(\Sigma)$ is better thought of as a geometric quantization of the phase space of solutions to the equations of motion on an infinitesimal strip $\Sigma\times[0,\epsilon]$ with a symplectic structure induced by $S$.} The physical meaning of this is that a state $\ket{\Psi}\in\mathcal{H}(\Sigma)$ assigns to a field configuration $\varphi\in\mathcal{E}_\Sigma$ the probability amplitude density $\ip{\varphi}{\Psi}$. However, the word \textbf{density} in here is key: this interpretation is only meaningful if one can construct a measure $D\varphi$ on $\mathcal{E}_\Sigma$. Then, the physical observables would correspond to the functions that allow to answer the question ``Given a measurable set of field configurations $E\subseteq\mathcal{E}_\Sigma$, what is the probability that the measurement of the field configuration would lie in $E$?," and the result will be\footnote{See \cite{Dowker2010} for a more thorough discussion of the appearance of this sort of integrals in quantum mechanics and their connection to the Schwinger-Keldysh formalism.}
\begin{equation}\label{eq:born-rule}
    \int_ED\varphi\, |\ip{\varphi}{\Psi}|^2.
\end{equation}

A general theory of such measures is beyond the scope of this paper. However, we expect the theory of $D\varphi$ to run along analogous lines as the theory of $\mathcal{D}\phi$ that we already discussed in the previous sections. Indeed, as we will see below, many of the integrals \eqref{eq:born-rule} are of the form
\begin{equation}
    \int_{\mathcal{E}_\Sigma}D\varphi\,e^{-S_\eff},
\end{equation}
with $S_\eff$ an effective action coming from a $\mathcal{D}\phi$ path integrals. Therefore, we can try to define these path integrals using the same technology we studied above. In particular, we will need to introduce an inner product on $\mathcal{E}_{\partial M}$, which will require the introduction of a scale $\nu$. By convention, we will take said scale with dimensions $[\nu]=[\mu]$. The resulting $D\varphi$ integral is, however, much more difficult to compute in general since the operator $Q_\partial$ for which
\begin{equation}
    S_\eff = \frac{1}{2}\ev{\varphi,\nu Q_\partial\varphi},
\end{equation}
is generically non-local. For example, the effective action will generically have some contribution from the instanton action, which depends in a non-local fashion on the incoming and outgoing boundary fields. 

However, it is useful to keep in mind the $D=1$ case, where QFT reduces to finite degrees of freedom quantum mechanics. In this case a dimensionless measure would have the form
\begin{equation}\label{eq:measure_D=1}
    D\varphi =\dd[N]{\varphi} \nu^{N[\phi]/[Q]}.
\end{equation}
One might be tempted at this stage to take $\nu=\mu$. As we argue below, though, $\mu$ is generally a function of $\nu$ of physical significance. It should also be noted that in topological field theories, one should also be able to construct the measure $D\varphi$ rigorously. Indeed, in these, the space of fields is heavily reduced by constraints and gauge freedom, making it effectively finite-dimensional.

The connection between the path integral formalism and the quantum theory lies in the equation\footnote{One might be worried about the meaning of $\ket{\varphi}$ in this equation. This equation is in the sense of distributions, meaning that
\begin{equation}
    \mel{\varphi_\Out}{U(M)}{\Psi}=\int_{\mathcal{E}_{\Sigma_\In}}D\varphi_{\In}\int_\mathcal{E}\mathcal{D}\phi\, e^{-S(\phi)}.
\end{equation}
When combined with \eqref{eq:normalization_hilbert}, using these improper vectors helps streamline the exposition.}
\begin{equation}\label{eq:quantum_interpretation}
    \mel{\varphi_\Out}{U(M)}{\varphi_\In}=\int_\mathcal{E}\mathcal{D}\phi\, e^{-S(\phi)},
\end{equation}
if the fields in $\mathcal{E}$ have the boundary conditions\footnote{In the mathematical literature \cite{Atiyah1988}, $M:\Sigma_\In\rightarrow\Sigma_\Out$ is interpreted as a cobordism. Then, setting $\mathcal{Z}(\Sigma)\equiv\mathcal{H}(\Sigma)$ and $\mathcal{Z}(M)\equiv U(M)$, the quantum theory is defined by the functor $\mathcal{Z}$ from cobordisms to vector spaces.}
\begin{equation}
    \phi|_{\Sigma_\In}=\varphi_\In \qand \phi|_{\Sigma_\Out}=\varphi_\Out.
\end{equation}
Here, $U(M)$ is the time evolution operator\footnote{If the action is real, this will be in Euclidean time.} and, after fixing $\mu$, we can use the path integral to define it. Thus, knowing the value of the transition amplitude on the left-hand side completely fixes $\mu$ on the right-hand side, and its value will depend on the measures of both the incoming and outgoing boundaries.

Since we have defined our path integrals to be dimensionless, equation \eqref{eq:quantum_interpretation} demands that the vectors $\ket{\varphi}$ are dimensionless for consistency. Additionally, the completeness relation
\begin{equation}\label{eq:normalization_hilbert}
    \int_{\mathcal{E}_\Sigma} D\varphi\, \op{\varphi} = 1.
\end{equation}
demands that the measure itself be dimensionless. Therefore, the introduction of $\nu$ is thus unavoidable. By not changing the scale between the incoming and outgoing boundaries, we aspire to fix the ratio $\nu/\mu$. So, this ratio has a physical meaning and is important to extract the predictions of the quantum theory at hand.

This still leaves the issue of obtaining a transition amplitude to compute $\mu$ using \eqref{eq:quantum_interpretation}. Fortunately, one can recover $\mu$ from consistency conditions any amplitude should satisfy. For instance, consider a spacetime of the form $M=[0, L]\times\Sigma$ and the limit $L\rightarrow 0^+$. Since the system has no time evolution, we must have $U(M)\rightarrow 1$ in this limit. We can then use the orthonormality condition
\begin{equation}
    \ip{\varphi_\Out}{\varphi_\In}=\delta(\varphi_\Out,\varphi_\In),
\end{equation}
as a proxy for the transition amplitude. Of course, $\delta$ is the Dirac delta function associated with the measure $D\varphi$ one chooses. This method works well for the bosonic $D=1$ case since, as we will see in \eqref{eq:1d_beta_line}, $\beta$ is a topological invariant. For $D=2$, however, $\beta\propto\Vol M$, so this method leaves a $\mu$-invariant path integral on the right-hand side of \eqref{eq:quantum_interpretation}. 

\begin{figure}[ht]
	\centering
	\begin{tikzpicture}
		\draw plot [smooth, tension = 1] coordinates {(0,0.5)(1,1)(2.3,0.8)(4,1.3)(4,0)(2.5,-0.7)(1,-0.5)(-1,-1)(0,0.5)};
        \draw (-1.04,-0.95) .. controls (-0.8,-1.5) and (-1, -2.5) .. (-0.85,-3.4);
	    \draw (4.33,0.7) .. controls (4,-1.5) and (4, -2.5) .. (4,-3.4);
        \draw plot [smooth, tension = 1] coordinates{(-0.5,-3.2)(1,-2.7)(2,-3)(3,-2.8)(4,-3.4)(3,-3.5)(2,-3.8)(1,-4)(-0.7,-3.6)(-0.5,-3.2)}; 
        \node at (2,0.2) {$\Sigma_\Out$};
        \node at (1,-3.3) {$\Sigma_\In$};
        \node at (-1.5,-2.7) {$M_1$};
        \node at (4.7,-1) {$M_2$};
        \draw (1.56,-1.9) ellipse (2.46 and 0.4);
        \node at (1.5,-1.9) {$\Sigma$};
    \end{tikzpicture}
	\caption{\label{fig:cobordism_composition}The spacetime $M$ of figure \ref{fig:cobordism} split by a codimension 1 surface $\Sigma$. This splits the spacetime into the spacetimes $M_1$ and $M_2$.}
\end{figure}

Once we are convinced of \eqref{eq:quantum_interpretation}, the completeness relation \eqref{eq:normalization_hilbert} also demands that we can cut and paste path integrals together. This way, we can create alternatives to \eqref{eq:quantum_interpretation} in which path integrals give both sides of the equation. The information of the scale $\nu$ is then completely contained in the gluing procedure. The simplest gluing one can do (see e. g. \citep{Bentov2021} for an application) is to slice $M$ along a codimension 1 surface $\Sigma$ so that it becomes the composition of $M_1$ and $M_2$, as shown in figure \ref{fig:cobordism_composition}. Then, using \eqref{eq:normalization_hilbert}, one can rewrite the operator equation
\begin{equation}\label{eq:naive_scheme_2}
    \mel{\varphi_\Out}{U(M)}{\varphi_\In}=\int_{\mathcal{E}_\Sigma} D\varphi\,\mel{\varphi_\Out}{U(M_2)}{\varphi}\mel{\varphi}{U(M_1)}{\varphi_\In}
\end{equation}
in path integral language 
\begin{equation}\label{eq:pasting}
    \int_{\mathcal{E}_{\varphi_\In\rightarrow\varphi_\Out}}\mathcal{D}\phi\,e^{-S}=\int_{\mathcal{E}_\Sigma}D\varphi\int_{\mathcal{E}_{\varphi\rightarrow\varphi_\Out}}\mathcal{D}\phi_2\,\int_{\mathcal{E}_{\varphi_\In\rightarrow\varphi}}\mathcal{D}\phi_1\,e^{-S(\phi_2)-S(\phi_1)}
\end{equation}
to obtain another consistency condition that allows one to fix $\mu$. For clarity, we have made the boundary conditions explicit in the equation above. 

\begin{figure}[ht]
	\centering
	\begin{tikzpicture}
		\draw (0,0) ellipse (2 and 0.7);
		\draw (-2,0) arc (180:360:2 and 2);
        \draw (-2,0) arc (180:0:2 and 3);
        \node at (0,-1.5) {$M_1$};
        \node at (0,1.5) {$M_2$};
        \node at (0,0) {$\Sigma$};
	\end{tikzpicture}
	\caption{\label{fig:inner_products}A closed manifold $M$ split by a codimension 1 surface $\Sigma$. Through this splitting, the path integral in this geometry can be interpreted as computing $\ip{M_1}{M_2}$.}
\end{figure}

\begin{figure}[ht]
	\centering
	\begin{tikzpicture}
		\draw (0,0) ellipse (2 and 0.7);
		\draw (-2,0) arc (180:360:2 and 2);
        \node at (0,-1.5) {$M$};
        \node at (0,0) {$\Sigma$};
	\end{tikzpicture}
	\caption{\label{fig:path_integral_states}A spacetime $M$ with a boundary $\Sigma$. The path integral on $M$ defines a state on $\Sigma$. Note that this is true regardless of whether $\Sigma$ is connected. For example, the path integral on the geometry of figure \ref{fig:cobordism} also defines a state on $\Sigma=\Sigma_\In\cup\Sigma_\Out$.}
\end{figure}

An interesting version of \eqref{eq:pasting} is obtained by taking $\Sigma_\In=\Sigma_\Out=\varnothing$, as depicted in figure \ref{fig:inner_products}. Indeed, in this case, $M$ is closed, and, as we will see, there are large classes of examples in which path integrals over closed manifolds are scale-independent. It is useful to interpret these constructions by noting that path integrals are mechanisms for creating states in the following sense. Define a state on $\Sigma$ by considering a manifold $M$ with boundary $\partial M=\Sigma$, such as the one shown in figure \ref{fig:path_integral_states}. Then, we have a state $\ket{M}\in\mathcal{H}(\Sigma)$, whose probability amplitude density for being in the configuration $\varphi\in\mathcal{E}_\Sigma$ is given by
\begin{equation}
    \ip{\varphi}{M}=\int_\mathcal{E}\mathcal{D}\phi\, e^{-S(\phi)},
\end{equation}
where the fields in $\mathcal{E}$ are equal to $\varphi$ at $\Sigma$. In this interpretation, the path integral of figure \ref{fig:inner_products} corresponds to the inner product between the vectors
\begin{equation}
    \ip{M_1}{M_2}=\int_\mathcal{E}\mathcal{D}\phi\, e^{-S(\phi)}.
\end{equation}
On the other hand, the operator $U(M)$ associated to figure \ref{fig:cobordism}, can be understood as a state in $\mathcal{H}(\Sigma_\Out)\otimes\mathcal{H}(\Sigma_\In)$.

\begin{figure}[ht]
	\centering
	\begin{tikzpicture}
        \draw (0,0) ellipse (2 and 3);
        \draw (0,-1) arc (-30:30:2);
        \draw (0.1,-0.75) arc (210:150:1.5);
        \draw (-1.05,0) ellipse (0.95 and 0.5);
        \node at (-1,0) {$\Sigma$};
        \node at (1,0) {$M$};
    \end{tikzpicture}
	\caption{\label{fig:splitting_torus}A spacetime $M$ cut by a codimension 1 submanifold $\Sigma$ in such a way that the resulting manifold $M_\text{cut}$ is connected.}
\end{figure}

Let us comment on a final useful way of cutting and pasting path integrals together. One can slice along a circle so that $M$ is not split into two disjoint submanifolds, as shown in figure \ref{fig:splitting_torus}. This leaves us with another equation that can be used to fix $\mu$
\begin{equation}
    \int_{\mathcal{E}_M}\mathcal{D}\phi\,e^{-S}=\int_{\mathcal{E}_\Sigma}D\varphi\int_{\mathcal{E}_{M_\text{cut}}}\mathcal{D}\phi\,e^{-S}.
\end{equation}
Applying \eqref{eq:quantum_interpretation} to $M_\text{cut}$, one can think of this as $\tr U(M_\text{cut})$.

At this point, it is useful to remark that these interpretations of the path integral are compatible with each other if one is willing to adopt the axiom\footnote{This is a standard axiom in functorial QFT \cite{Atiyah1988}.} that the empty set supports a Hilbert space of dimension 1, i. e., $\mathcal{H}(\varnothing)=\Span\{\ket{\varnothing}\}\cong\mathbb{C}$. This axiom is natural from the point of view that any field on $\varnothing$ is a function whose domain is $\varnothing$. Thinking of functions as relations on the Cartesian product of their domain and their codomain, it is clear that there is only one such configuration, the empty configuration $\varphi=\varnothing$. Accordingly, the Hilbert space is spanned by a state that corresponds to this configuration and can be taken to be normalized
\begin{equation}
    \ip{\varnothing}=1.
\end{equation}
One can always decompose the boundary of a manifold so that the incoming boundary is $\varnothing$ and the outgoing boundary is $\Sigma=\partial M$. Then, the path integral defines a time evolution operator $U(M):\mathcal{H}(\varnothing)\rightarrow\mathcal{H}(\Sigma)$ which, due to linearity, is completely defined by the state $\ket{M}=U(M)\ket{\varnothing}$. From this point of view, $\ket{M}$ is the state obtained by evolving $\ket{\varnothing}$ along $M$. With the interpretation of the state $\ket{\varnothing}$, the two interpretations of the state $\ket{M}$ we have presented coincide. Furthermore, if $\partial M=\varnothing$, the operator $U(M)$ is completely determined by the expectation value $\ev{U(M)}{\varnothing}$.   

\section{Leveraging the Scale Dependences of \texorpdfstring{$D\varphi$}{D varphi} and \texorpdfstring{$\mathcal{D}\phi$}{D phi}}\label{sec:leveraging}

Having provided several frameworks in which one could fix $\mu$, let us now understand how the value of $\mu$ changes as we change the incoming and outgoing boundary measures. Such changes happen, for example, if we change the scale $\nu$. Assuming we have a second measure $\widetilde{D\varphi}$, absolutely continuous with respect to $D\varphi$, the completeness relation \eqref{eq:normalization_hilbert} guarantees that the basis vectors transform as
\begin{equation}
    \ket{\varphi,\widetilde{D\varphi}}=\sqrt{\frac{D\varphi}{\widetilde{D\varphi}}}\ket{\varphi,D\varphi},
\end{equation}
where we have made the dependence of $\ket{\varphi}$ on the measure explicit. Using \eqref{eq:multiplicative_constant}, we then conclude that in the setting of \eqref{eq:quantum_interpretation}
\begin{equation}
\begin{aligned}
    &\hphantom{=}\qty(\frac{\mu(D\varphi)}{\mu(\widetilde{D\varphi})})^{\beta+\dim\mathcal{J}/2}\int_\mathcal{E}(\mathcal{D}\phi)_{\mu(D\varphi)}\, e^{-S(\phi)}=\int_\mathcal{E}(\mathcal{D}\phi)_{\mu(\widetilde{D\varphi})}\, e^{-S(\phi)}\\
    &=\mel{\varphi_\Out,\widetilde{D\varphi}}{U(M)}{\varphi_\In,\widetilde{D\varphi}}=\sqrt{\frac{D\varphi}{\widetilde{D\varphi}}(\varphi_\Out)\frac{D\varphi}{\widetilde{D\varphi}}(\varphi_\In)}\mel{\varphi_\Out,D\varphi}{U(M)}{\varphi_\In,D\varphi}.
\end{aligned}
\end{equation}
To keep the notation as clean as possible, we have denoted the measures on $\Sigma_{\In}$ and $\Sigma_{\Out}$ by the same symbols. It is important to remember that they are generally different (and must be if $\Sigma_\In\neq\Sigma_\Out$). More precisely, $D\varphi$ in here corresponds to a measure on $\mathcal{E}_{\partial M}$. We will assume that this whole measure depends on a single scale $\nu$. We thus obtain the result
\begin{equation}\label{eq:change_measure}
    \qty(\frac{\mu(D\varphi)}{\mu(\widetilde{D\varphi})})^{\beta+\dim\mathcal{J}}=\sqrt{\frac{D\varphi}{\widetilde{D\varphi}}(\varphi_\Out)\frac{D\varphi}{\widetilde{D\varphi}}(\varphi_\In)}.
\end{equation}
Other scaling results can be obtained by applying this same logic to the settings of figures \ref{fig:cobordism_composition}, \ref{fig:inner_products}, and \ref{fig:splitting_torus}.

Fully understanding the implications of \eqref{eq:change_measure} would require a full theory of measures on the boundary conditions of a path integral for $D>1$. However, we can use this relationship to constrain what such a theory would imply. First, let us note that even if one had measures on $\mathcal{E}_{\partial M}$ for which $\mu$ is independent of the boundary conditions, a non-constant scaling of such a measure would introduce a dependence of $\mu$ on such boundary conditions. This selects a privileged measure up to rescaling, for which $\mu$ is independent of the boundary conditions. Conversely, this connects the inner product on $M$ and its codimension 1 submanifold $\Sigma$.

It is possible that further exploration of the scaling relations analog to \eqref{eq:change_measure} may be enough to fix $\beta$ completely. To see this, let us now show this for a $D=1$ bosonic theory. Let us assume that there exists a dimension $P$ for which
\begin{equation}\label{eq:scaling_submeasure}
    (D\varphi)_{\tilde{\nu}}=\qty(\frac{\tilde{\nu}}{\nu})^{P}(D\varphi)_\nu.\footnote{Then $-[\nu]P$ would be the dimension of a scale-independent measure on $\mathcal{E}_{\partial M}$.}
\end{equation}
We then have
\begin{equation}
    \qty(\frac{\mu}{\tilde{\mu}})^{\beta+\dim\mathcal{J}}=\qty(\frac{\nu}{\tilde{\nu}})^{P}
\end{equation}
Then, if $\mu$ is proportional to $\nu$, this equation is only consistent if
\begin{equation}\label{eq:index}
    P=\beta+\dim\mathcal{J}/2=\beta+\dim\ker Q/2.
\end{equation}
This provides an index-like formula for the beta function in $D=1$. Indeed, given \eqref{eq:measure_D=1}, we have
\begin{equation}
    \beta =\frac{N[\phi]}{[Q]}-\dim\ker Q/2=-N/4-\dim\ker Q/2
\end{equation}
This is verified in the explicit harmonic oscillator example in \eqref{eq:1d_beta_line}.

Let us finalize this section by providing a family of self-normalizing path integrals. In most applications, the definition of $\ev{\cdot,\cdot}$ requires the introduction of a metric on spacetime. In fact, the bilinear form will be proportional to powers of the metric. Accordingly, the factor $\mu$ can equivalently be understood as rescaling the metric. Other than the implications this might have for the metric structure of spacetime at the quantum level, this point of view is particularly interesting in the setting of topological field theories. At the classical level, topological field theories can often be described by actions that are manifestly independent of the metric. However, introducing the bilinear form required to understand their path integral demands that one equip spacetime with a fiduciary metric. If the results of one's computation depend on the metric, then the theory is no longer topological at the quantum level. In particular, any non-anomalous topological field theory must also be self-normalizing. 

\section{Examples}\label{sec:examples}

\subsection{Scalars}

Let us begin with the important example of the Klein-Gordon field $\phi\in C^\infty(M)$ on a Riemannian manifold $M$ with action
\begin{equation}\label{eq:scalar_action}
    S(\phi)=\int_M\dd[D]{x}\sqrt{g}\qty(\frac{1}{2}g^{\mu\nu}\partial_\mu\phi\partial_\nu\phi+\frac{1}{2}m^2\phi^2).
\end{equation}
In $D=1$, we should recover the familiar results for the partition function of the harmonic oscillator. In $D=2$, this theory is part of the bosonic string.

\subsubsection{General Closed Manifolds}\label{sec:scalar_closed}

Let us begin by assuming that $M$ is closed. Then, the path integral we want to compute is
\begin{equation}
    \int_{C^\infty(M)}\mathcal{D}\phi\,e^{-S(\phi)}.
\end{equation}
We can now introduce the inner product
\begin{equation}\label{eq:inner_product_Klein}
    \ev{\phi,\psi}=\frac{1}{2\pi\mu}\int\dd[D]{x}\sqrt{g}\phi\psi.
\end{equation}
Integrating the action by parts (and noticing this induces no boundary terms since $\partial M=\varnothing$), we also obtain the differential operator
\begin{equation}
    Q=-\Delta+m^2,
\end{equation}
so that
\begin{equation}
    S(\phi)=\frac{1}{2}\ev{\phi,2\pi \mu Q\phi}.
\end{equation}
On closed manifolds, the Laplacian $-\Delta$ is non-negative (as shown by integration by parts), with the eigenvalue $0$ having multiplicity 1. In particular, for $m^2>0$, we have a positive operator, and we can conclude that
\begin{equation}
    \int_{C^\infty(M)}\mathcal{D}\phi\,e^{-\frac{1}{\hbar}S(\phi)}=e^{-S_{\text{eff}}(\mu)}\qc S_{\text{eff}}(\mu)=-\frac{1}{2}\zeta'_{\mu Q}(0)
\end{equation}

The beta function is then obtained from the heat kernel of this operator, which has the asymptotic expansion \citep{lablee:hal-01504945}
\begin{equation}\label{eq:kernel_closed}
\begin{aligned}
    \tr e^{-t\mu Q}=e^{-t\mu m^2}\tr(e^{t\mu \Delta})\sim\frac{e^{-t\mu m^2}}{(4\pi\mu  t)^{D/2}}\qty(a_0+a_1\mu t+a_2(\mu t)^2+\cdots),
\end{aligned}
\end{equation}
whose coefficients are integrals over the metric and its derivatives, e.g. $a_0=\Vol M$. We then immediately conclude that, for odd $D$, we have $\beta=0$. For even $D$, we have
\begin{equation}
    \beta=\frac{1}{2(4\pi)^{D/2}}\sum_{k=0}^{D/2}\frac{(-1)^km^{2k}}{k!}b_{D/2-k}
\end{equation}
Thus, for example, in $D=2$ we have $b_1 = \pi\chi(M)/3$, i.e.
\begin{equation}\label{eq:2d_beta}
    \beta=\frac{\chi(M)}{24}-\frac{m^2\Vol M}{8\pi},
\end{equation}
where $\chi$ is the Euler characteristic.

\subsubsection{The \texorpdfstring{$D$}{D}-Torus}\label{app:scalar_torus}

Let us consider a $D$-torus
\begin{equation}
    T^D=\underbrace{S^1\times\cdots\times S^1}_{D\text{ times}}.
\end{equation}
On a product manifold, the Laplacian factorizes into a sum of the Laplacian on each factor. The heat kernel, in turn, factorizes into the tensor product of the individual heat kernels so that their traces get multiplied. Therefore, understanding the zeta function through the Mellin transform \eqref{eq:mellin_transform} requires one to understand the zeta function on each factor.

The zeta function on the circle of circumference $L$ is obtained by noting that the eigenfunctions of its Laplacian are periodic plane waves so that
\begin{equation}\label{eq:kernel_circle}
    e^{t\mu\Delta}=\sum_{k\in\frac{2\pi}{L}\mathbb{Z}}e^{-t\mu k^2}\op{k}.
\end{equation}
We can isolate the $t\rightarrow 0^+$ divergence by using Poisson resummation. This is obtained by localizing the analogous Gaussian integral to the sum by using a periodic delta function
\begin{equation}
    e^{t\mu\Delta}=\int\dd{p}\qty(\sum_{k\in\frac{2\pi}{L}\mathbb{Z}}\delta(k-p))e^{-t\mu p^2}\op{p}.
\end{equation}
This periodic delta function can be thought of as the wave function of a particle sitting at a definite position $p=0$ on a circle of circumference $2\pi / L$. Going to momentum space on said circle, we have plane waves
\begin{equation}
    \ip{p}{l}=\sqrt{\frac{L}{2\pi}}e^{ilp}\qc l\in L\mathbb{Z}.
\end{equation}
We then have
\begin{equation}
    \sum_{k\in\frac{2\pi}{L}\mathbb{Z}}\delta(k-p)=\ip{p}{0}=\sum_{l\in L\mathbb{Z}}\ip{p}{l}\ip{l}{0}=\frac{L}{2\pi}\sum_{l\in L\mathbb{Z}}e^{ilp},
\end{equation}
with which we can eliminate the delta function 
\begin{equation}\label{eq:poisson_resummation}
    e^{t\mu\Delta}=L\sum_{l\in L\mathbb{Z}}\int\frac{\dd{p}}{2\pi}e^{-t\mu p^2+ilp}\op{p}.
\end{equation}
To compute its trace, we can perform the remaining Gaussian integral
\begin{equation}
    \tr(e^{t\mu\Delta})=L\sum_{l\in L\mathbb{Z}}\int\frac{\dd{p}}{2\pi}e^{-t\mu p^2+ilp}=\frac{L}{\sqrt{4\pi\mu t}}\sum_{l\in L\mathbb{Z}}e^{-\frac{l^2}{4\mu t}}.
\end{equation}
The $t\rightarrow 0^+$ divergence has been isolated to the prefactor in this form. 

Going back to the $D$-torus, we have
\begin{equation}\label{eq:zeta_torus}
\begin{aligned}
    \zeta_{\mu Q}(s)&=\frac{\Vol M}{\Gamma(s)(4\pi\mu )^{D/2}}\sum_{\vec{l}\in\Lambda}\int_0^\infty\dd{t}t^{s-1-D/2}e^{-t\mu m^2-\frac{\vec{l}^2}{4\mu t}}\\
    &=\frac{\Vol M}{\Gamma(s)(4\pi\mu )^{D/2}}\sum_{\vec{l}\in\Lambda}\begin{cases}
    2\qty(\frac{4\mu^2m^2}{\vec{l}^2})^{\frac{D/2-s}{2}}K_{D/2-s}(m\|\vec{l}\|) & \vec{l}\neq \vec{0},\\
    (\mu m^2)^{D/2-s}\Gamma(s-D/2) & \vec{l} = \vec{0}.
    \end{cases}
\end{aligned}
\end{equation}
with $\Lambda$ the lattice
\begin{equation}
    \Lambda = L_1\mathbb{Z}\times\cdots\times L_n\mathbb{Z},
\end{equation}
where $L_1,\dots,L_n$ are the circumferences of the circles. The $s\rightarrow 0$ limit is simplified by noting that
\begin{equation}
    \lim_{s\rightarrow 0}\frac{K_{D/2-s}(m\|\vec{l}\|)}{\Gamma(s)}=0\qc \vec{l}\neq \vec{0},
\end{equation}
and
\begin{equation}
    \lim_{s\rightarrow 0}\frac{\Gamma(s-D/2)}{\Gamma(s)}=\begin{cases}
        0 & D\text{ odd,}\\
        \lim_{s\rightarrow 0}\frac{1}{(s-1)(s-2)\cdots (s-D/2)}=\frac{(-1)^{D/2}}{(D/2)!} & D\text{ even}
    \end{cases}
\end{equation}
so that
\begin{equation}
    \zeta_{\mu Q}(0)=\begin{cases}
        0 & D\text{ odd},\\
        (-1)^{D/2}\frac{m^D\Vol M}{(4\pi)^{D/2}(D/2)!} & D\text{ even}.
    \end{cases}
\end{equation}
Comparison with \eqref{eq:2d_beta} allows us to conclude that $b_n=0$ for $n\in\{1,\dots, D/2\}$ in the $D$-torus for $D$ even.

We can now calculate the effective action by taking the derivative of the zeta function
\begin{equation}
\begin{aligned}
    \zeta'_{\mu Q}(0)&=\frac{\Vol M}{(4\pi \mu)^{D/2}}\left(-(\mu m^2)^{D/2}\ln(\mu m^2)\lim_{s\rightarrow 0}\frac{\Gamma(s-D/2)}{\Gamma(s)}\right.\\
    &+(\mu m^2)^{D/2}\left.\dv{s}\frac{\Gamma(s-D/2)}{\Gamma(s)}\right|_{s=0}\\
    &\left.+2\sum_{\vec{l}\in\Lambda\setminus\{\vec{0}\}}\qty(\frac{4\mu^2m^2}{\|\vec{l}\|^2})^{D/4}\left.\dv{s}\frac{K_{D/2-s}(m\|\vec{l}\|)}{\Gamma(s)}\right|_{s=0}\right).
\end{aligned}
\end{equation}
Now, since
\begin{equation}
    \left.\dv{s}\frac{K_{D/2-s}(m\|\vec{l}\|)}{\Gamma(s)}\right|_{s=0}=K_{D/2}(m\|\vec{l}\|),
\end{equation}
for any $D$, we have that
\begin{equation}
\begin{aligned}
    \zeta'_{\mu Q}(0)&=\Vol M\left(\frac{m^2}{4\pi}\right)^{D/2}\left(-\ln(\mu m^2)\lim_{s\rightarrow 0}\frac{\Gamma(s-D/2)}{\Gamma(s)}\right.\\
    &+\left.\dv{s}\frac{\Gamma(s-D/2)}{\Gamma(s)}\right|_{s=0}\\
    &\left.+2^{1+D/2}\sum_{\vec{l}\in\Lambda\setminus\{\vec{0}\}}\qty(\frac{1}{m\|\vec{l}\|})^{D/2}K_{D/2}(m\|\vec{l}\|)\right).
\end{aligned}
\end{equation}

For odd $D$, 
\begin{equation}
    \zeta'_{\mu Q}(0)=\Vol M\left(\frac{m^2}{4\pi}\right)^{D/2}\left(\Gamma(-D/2)+2^{1+D/2}\sum_{\vec{l}\in\Lambda\setminus\{\vec{0}\}}\qty(\frac{1}{m\|\vec{l}\|})^{D/2}K_{D/2}(m\|\vec{l}\|)\right)
\end{equation}
In particular, if $D=1$, the Bessel function $K_{1/2}(x)$ can be expressed in such a way that we obtain a closed-form solution for the path integral \citep{arfken_mathematical_2012, gradshteyn_table_2014}
\begin{equation}\label{eq:zeta_circle}
    \zeta'_{\mu Q}(0)=mL\left(-1+\frac{1}{mL}\sum_{n\in\mathbb{Z}\setminus{\{0\}}}\frac{e^{-mL|n|}}{|n|}\right)=-mL-2\ln(1-e^{-mL}),
\end{equation}
which leads us to the effective action
\begin{equation}
    S_\text{eff}=\frac{1}{2}mL+\ln(1-e^{-mL})=\ln(2\sinh(mL/2))
\end{equation}
which coincides with the partition function of the harmonic oscillator.

In contrast, for even $D$ we have
\begin{equation}
\begin{aligned}
    \zeta'_{\mu Q}(0)&=\Vol M\left(\frac{m^2}{4\pi}\right)^{D/2}\left(-\frac{(-1)^{D/2}}{(D/2)!}\ln(\mu m^2)\right.\\
    &+\left.\dv{s}\frac{1}{(s-1)(s-2)\cdots (s-D/2)}\right|_{s=0}\\
    &\left.+2^{1+D/2}\sum_{\vec{l}\in\Lambda\setminus\{\vec{0}\}}\qty(\frac{1}{m\|\vec{l}\|})^{D/2}K_{D/2}(m\|\vec{l}\|)\right).
\end{aligned}
\end{equation}
Even if an explicit calculation of the path integral becomes more intractable as we go up in the dimension of our manifold, we can rest assured that the series will converge since $\int_{a}^{\infty}x^{-n} K_n(x)\dd x$ is finite for any $a$ and $n$ and since $\lim_{x\rightarrow 0}x^{-n} K_n(x)=0$ for any $n$.

\subsubsection{General Cylinders}\label{sec:scalar_cylinder}

Let us now consider a manifold of the form $M=[0,L]\times\Sigma$, with $\Sigma$ a closed manifold. This manifold has incoming $\Sigma_\In=\Sigma\times\{0\}$ and outgoing boundaries $\Sigma_\Out=\Sigma\times\{L\}$. Thus, given $\varphi_\In\in C^\infty(\Sigma_\In)$ and $\varphi_\Out\in C^\infty(\Sigma_\Out)$, we can define the space of fields
\begin{equation}
    \mathcal{E}=\set{\phi\in C^\infty(M)|\phi|_{\Sigma_\In}=\varphi_\In\qand \phi|_{\Sigma_\Out}=\varphi_\Out},
\end{equation}
so that the path integral computes the transition amplitude in Euclidean time
\begin{equation}
    \begin{tikzpicture}[baseline]
    \draw (0,0.5) ellipse (0.5 and 0.25);
    \draw (0.5,-0.5) arc (0:-180:0.5 and 0.25);
    \draw[dashed] (0.5,-0.5) arc (0:180:0.5 and 0.25);
    \draw (-0.5,-0.5) -- (-0.5,0.5);
    \draw (0.5,-0.5) -- (0.5,0.5);
    \node at (0,1) {$\varphi_\Out$};
    \node at (0,-1) {$\varphi_\In$};
    \end{tikzpicture}\equiv\int_\mathcal{E}\mathcal{D}\phi\,e^{-S(\phi)}=\mel{\varphi_\Out}{U(M)}{\varphi_\In}.
\end{equation}

In this case, $\mathcal{E}$ is not a vector space. We must, therefore, study the space of instantons of this theory, which satisfy the equations of motion $\Delta\phi=m^2\phi$. For these, the action becomes
\begin{equation}\label{eq:instanton_action}
    S(\phi)=\int_{\partial M}\dd[D-1]{x} n^\mu\phi\partial_\mu\phi,
\end{equation}
after integration by parts. This, in particular, shows the instanton is unique. Indeed, the difference between two instantons is also an instanton but with vanishing boundary conditions on $\partial M$. Therefore, its action vanishes. However, from the bulk form of the action, it is clear that this implies that the difference vanishes. Let us then denote the unique instanton by $\phi_c$.

The vector bundle is then trivial, with a base $\mathcal{J}=\{\phi_c\}$. The projection is uniquely defined, and the space of quantum fluctuations is 
\begin{equation}
    \mathcal{F}=\set{\phi\in C^\infty(M)|\phi|_{\partial M}=0},
\end{equation}
so that our space of fields is affine
\begin{equation}
    \mathcal{E}=\phi_c+\mathcal{F}.
\end{equation}
The effective action is then
\begin{equation}
    S_\eff=S(\phi_c)-\frac{1}{2}\zeta'_{\mu Q}(0)
\end{equation}

Due to the vanishing boundary conditions, we can integrate by parts on $\mathcal{F}$. Accordingly, $S_\eff$ is built from the zeta function of $\mu Q$ with $Q=-\Delta+m^2$. The Laplacian factors into the sum of the one on $\Sigma$ and the one on the interval 
$[0,L]$. Accordingly, its associated heat kernel becomes the tensor product of the heat kernel on each factor. For the heat kernel on $\Sigma$, we have the result \eqref{eq:kernel_closed}. For the one on the interval, the eigenfunctions are given by sine waves vanishing at the boundaries, so that\footnote{Note the difference in the summation range when compared to \eqref{eq:kernel_circle}. This is because the eigenspaces for $k\neq 0$ in this case are non-degenerate, and for $k=0$, we no longer have a non-vanishing eigenconfiguration.}
\begin{equation}\label{eq:heat_line}
    \tr(e^{t\mu \Delta_{[0,L]}})=\sum_{k\in \frac{\pi}{L}\mathbb{N}^+}e^{-t\mu k^2}=\frac{1}{2}\qty(\sum_{k\in \frac{\pi}{L}\mathbb{Z}}e^{-t\mu k^2}-1)=\frac{1}{2}\qty(\tr(e^{t\mu\Delta_{S^1}})-1),
\end{equation}
with $S^1$ a circle of circumference $2L$.\footnote{The main difference between the computation of the heat kernel in this circle and the corresponding interval is that on the circle, each eigenvalue $k\neq 0$ is doubly degenerate since both sine and cosine waves are admissible. Furthermore, the sine wave vanishes identically in the zero mode $k=0$.} The $t\rightarrow 0^+$ divergence can be isolated by using Poisson resummation to obtain
\begin{equation}
    \tr(e^{t\mu \Delta_{[0,L]}})=\frac{L}{\sqrt{4\pi\mu t}}\sum_{l\in 2L\mathbb{Z}}e^{-\frac{l^2}{4\mu t}} - \frac{1}{2}\sim\frac{L}{\sqrt{4\pi\mu t}}-\frac{1}{2}.
\end{equation}

This allows us to read the asymptotic expansion for the full heat kernel in terms of the one for $\Sigma$
\begin{equation}
\begin{aligned}
    \tr(e^{t\mu\Delta})&=\tr(e^{t\mu\Delta_\Sigma})\tr(e^{t\mu\Delta_{[0,L]}})\\
    &\sim\frac{1}{(4\pi\mu t)^{(D-1)/2}}\qty(a_0 + a_1\mu t+a_2(\mu t)^2+\cdots)\qty(\frac{L}{\sqrt{4\pi\mu t}}-\frac{1}{2})
\end{aligned}
\end{equation}
We conclude that
\begin{equation}
    \zeta_Q(0)=\begin{cases}\frac{L}{(4\pi)^{D/2}}\sum_{k=0}^{D/2}\frac{(-1)^km^{2k}}{k!}a_{D/2-k}& D\text{ even}\\
    -\frac{1}{2(4\pi)^{(D-1)/2}}\sum_{k=0}^{(D-1)/2}\frac{(-1)^km^{2k}}{k!}a_{(D-1)/2-k }& D\text{ odd}.
    \end{cases}
\end{equation}
For example, for $D=1$, we have
\begin{equation}\label{eq:1d_beta_line}
    \beta=-\frac{1}{4},
\end{equation}
in agreement with \eqref{eq:index}. For $D=2$ we have
\begin{equation}\label{eq:beta_cylinder}
    \beta = -\frac{m^2\Vol M}{8\pi},
\end{equation}
since for $\Sigma=S^1$ we have $a_1=0$.

\subsubsection{Cylinder with Toric Lids \texorpdfstring{$[0, L]\times T^{D-1}$}{[0,L] x T(D-1)}}

If our manifold is a cylinder with toric lids, the heat kernel factorizes between one for the line $[0, L]$ and the $D-1$-torus $T^{D-1}$, both of which we have already studied. Remembering that the heat kernel of a line \eqref{eq:heat_line} is related to that of a circle $S^1$ of circumference $2L$, we have the following result for the case we are interested in:
\begin{equation}
    \zeta_{[0, L]\times T^{D-1}}(s)=\frac{1}{2}\qty(\zeta_{T^D_{2L,L_1,\dots,L_{D-1}}}(s)-\zeta_{T^{D-1}}(s))
\end{equation}
Then, for concrete cases, we can simply take the results from our last section and bear the parity of $D$ in mind. For instance, for $D=1$,
\begin{equation}\label{eq:line}
    \zeta'_{[0, L]}(0)=\frac{1}{2}(-2mL-2\ln(1-e^{-2mL})+\ln{\mu m^2})=\ln(\frac{\sqrt{\mu m^2}}{2\sinh(mL)}),
\end{equation}
where we have used that since the 0-torus is a point, its volume is 1 by convention, and there is no sum over a lattice of values on its circumference. If we choose $\varphi_{\text{in}}$ to be the unique instanton determined by the boundary conditions and the equations of motion, the classical action is given by
\begin{equation}
    S(\phi_c)=\frac{m}{2}\qty(\qty(\varphi_{\text{out}}^2+\varphi_{\text{in}}^2)\coth(mL)-2\varphi_{\text{out}}\varphi_{\text{in}}\csch(mL)).
\end{equation}
Using \eqref{eq:line}, we conclude that:
\begin{equation}\label{eq:transition_harmonic_oscillator}
\int\mathcal{D}\phi\,e^{-S}=\mu^{1/4}\sqrt{\frac{m}{2\sinh{mL}}}e^{-\frac{m}{2}\qty(\qty(\varphi_{\text{out}}^2+\varphi_{\text{in}}^2)\coth(mL)-2\varphi_{\text{out}}\varphi_{\text{in}}\csch(mL))}
\end{equation}

To finish this calculation, we determine $\mu$ by using the cut-paste condition that joining the two ends of a line should yield a circle and equation \eqref{eq:measure_D=1}
\begin{equation}
\begin{aligned}
    \frac{1}{2\sinh(mL/2)}&=\mu^{1/4}\sqrt{\frac{m}{2\sinh{mL}}}\int\frac{\dd{\varphi}}{\nu^{1/4}}e^{-m\qty(\coth(mL)-\csch(mL))\varphi^2}\\
    &=\left(\frac{\mu}{\nu}\right)^{1/4}\frac{\sqrt{\pi}}{2\sinh(mL/2)}.
\end{aligned}
\end{equation}
This fixes
\begin{equation}\label{eq:renormalization_constant}
    \frac{\mu}{\nu}=\frac{1}{\pi^2}.
\end{equation}

\subsubsection{Scale comparison on \texorpdfstring{$[0,L]\times S^1$}{[0,L] x S}}\label{sec:scale_D_2}

Let us now focus on the $D=2$ case of the previous paragraph. We will argue that the relationship \eqref{eq:renormalization_constant} remains true in this case. We will follow a logic suggested to us by Bruno de S. L. Torres \cite{bruno2023}. Although we are in $D=2$, this computation is tractable because we can reduce this theory to an infinite set of uncoupled $D=1$ by thinking of this as a quantum mechanical theory with infinite degrees of freedom
\begin{equation}
    C^\infty([0,L]\times S^1)=C^\infty([0,L],C^\infty(S^1))
\end{equation}
Indeed, we can introduce a dimensionless inner product on the space of fields at a constant time slice $C^\infty(S^1)$
\begin{equation}\label{eq:inner_product_lids}
    \ev{\varphi_1,\varphi_2}_{S^1}=\int_{S^1_{L_1}}\frac{\dd{x}}{\sqrt{\nu}}\,\varphi_1(x)\varphi_2(x),
\end{equation}
with $\nu$ a scale of dimensions $[\nu]=[\mu]=-2$. To make the infinite degrees of freedom manifest, let us introduce the orthonormal basis of eigenfields of the Laplacian on the circle, given by
\begin{equation}
    \varphi_n(x)=\sqrt{\frac{\sqrt{\nu}}{L_1}}e^{ik_nx}\qc k=\frac{2\pi n}{L_1}\qc n\in\mathbb{Z}
\end{equation}
Expanding our fields in this basis
\begin{equation}
    \phi(t,x)=\sum_{n\in\mathbb{Z}}\phi^n(t)\varphi_n(x)\qc(\phi^n)^*=\phi^{-n},
\end{equation}
the action becomes
\begin{equation}
    S=\sum_{n\in\mathbb{Z}}S_{[0,L]}(\phi^n)=\sum_{n\in\mathbb{Z}}\int\dd{t}\sqrt{\nu}\qty(\frac{1}{2}|\dot{\phi^n}|^2+\frac{1}{2}m_n^2|\phi^n|^2)\qc m_n^2=m^2+k_n^2.
\end{equation}

The $D=1$ path integrals are computed through a dimensionless inner product on the interval
\begin{equation}
    \ev{\phi_1,\phi_2}_{[0,L]}=\frac{1}{2\pi\mu}\int\dd{t}\sqrt{\nu}\phi_1(t)\phi_2(t).
\end{equation}
We have decided to use the same bulk scale to make this inner product adimensional so that the bulk inner product takes the form
\begin{equation}
\begin{aligned}
    \ev{\phi_1,\phi_2}&=\frac{1}{2\pi\mu}\int\dd{t}\sqrt{\nu}\ev{\phi_1(t,\cdot),\phi_2(t,\cdot)}_{S^1}=\frac{1}{2\pi\mu}\sum_{n\in\mathbb{Z}}\int\dd{t}\sqrt{\nu}\phi_1^n(t)^*\phi_2^n(t)\\
    &=\sum_{n\in\mathbb{Z}}\ev{\phi_1^n,\phi_2^n}.
\end{aligned}
\end{equation}
This choice gives us
\begin{equation}
    \mathcal{D}\phi=\prod_{n\in\mathbb{Z}}\mathcal{D}\phi^n,
\end{equation}
and we have that the bulk path integral reduces to a product of $D=1$ path integrals
\begin{equation}\label{eq:product}
    \int\mathcal{D}\phi\, e^{-S}=\prod_{n\in\mathbb{Z}}\int\mathcal{D}\phi^n\,e^{-S_{[0,L]}(\phi^n)}.
\end{equation}

In order to compute these path integrals, we can use the result from the previous section. Note that the scale $\nu$ only appears on the instanton contributions of these integrals
\begin{equation}
    S_{[0,L]}(\phi^n_c)=\frac{\sqrt{\nu}m}{2}\qty(\qty((\varphi^n_{\text{out}})^2+(\varphi^n_{\text{in}})^2)\coth(m_nL)-2\varphi^n_{\text{out}}\varphi^n_{\text{in}}\csch(m_nL)),
\end{equation}
where we decomposed
\begin{equation}
    \varphi_{\In/\Out}(x)=\sum_{n\in\mathbb{Z}}\varphi_{\In/\Out}^n\varphi_n(x)
\end{equation}
This was to be expected since the $\nu$ scale should only appear at the boundaries of the theory. Once this contribution is dealt with, the remaining integral has vanishing boundary conditions. Therefore, we can integrate by parts and express the action in terms of the inner product in the form
\begin{equation}
    S_{[0,L]}(\phi^n)=\frac{1}{2}\ev{\phi^n,2\pi\mu Q_n\phi^n}_{[0,L]}\qc Q_n = -\Delta_{[0,L]}+m_n^2.
\end{equation}
We conclude that the path integral is 
\begin{equation}
    \int\mathcal{D}\phi_n\, e^{-S_{[0,L]}(\phi_n)}=\mu^{1/4}\sqrt{\frac{m_n}{2\sinh(m_n L)}}e^{-S_{[0,L]}(\phi_c)}.
\end{equation}

Computing the remaining product over modes in \eqref{eq:product} would require yet another zeta function regularization. This is outside of the scope of this paper since the resulting zeta function is that of a non-local theory, as is expected since the two circles comprising the boundary of our spacetime are disconnected from one another. This is reflected in the fact that the remaining terms in the product are not polynomials in $m_n$. However, one can bravely compute the relationship between $\mu$ and $\nu$ by relating this result to the result on the torus obtained by pasting the two boundaries of the cylinder together. Indeed, this path integral can also be reduced to a path integral over modes following the steps outlined above, with each mode corresponding to a path integral on the temporal circle. The resulting path integral is, however, independent of $\nu$ since the space of fields is a vector space and no instanton contributions are required. In fact, it is also independent of $\mu$ since, as we already showed, scalar path integrals on odd-dimensional spaces self-normalize. The contribution to each mode is therefore 
\begin{equation}
    \frac{1}{2\sinh(m_n L/2)}.
\end{equation}

In order to relate these two results, we use the boundary measure induced by the inner product \eqref{eq:inner_product_lids}. This corresponds to integrating over the $\varphi^n$ coordinates. Thus, we must have
\begin{equation}
\begin{aligned}
    \frac{1}{2\sinh(m_nL/2)}&=\mu^{1/4}\sqrt{\frac{m_n}{2\sinh(m_n L)}}\int\dd{\varphi^n}e^{-\sqrt{\nu}m_n(\varphi^n)^2(\coth(m_nL)-\csch(m_nL))}\\
    &=\qty(\frac{\mu}{\nu})^{1/4}\frac{\sqrt{\pi}}{2\sinh(m_nL/2)}.
\end{aligned}
\end{equation}
We conclude that
\begin{equation}
    \frac{\mu}{\nu}=\frac{1}{\pi^2},
\end{equation}
just like in the $D=1$ theory.

\subsection{Compact Boson}

Let us now consider a scalar field $\phi\in C^\infty(M, S^1)$ valued in a circle $S^1$ of circumference $\ell$. Let the action still be given by \eqref{eq:scalar_action} with $m^2=0$.\footnote{Since $\phi^2$ is not a well-defined function on the circle, this is the only quadratic theory possible with this field content.} Even for closed $M$, the space of fields is not a vector space since $S^1$ does not have a vector space structure. One can try to amend this by thinking of $\phi$ as a complex-valued field and writing
\begin{equation}
    \phi=e^{2\pi i\alpha/\ell},
\end{equation}
for a real-valued field $\alpha$ on $M$. Mathematically, $\alpha$ is a lift of $\phi$ to the universal cover $\mathbb{R}\rightarrow S^1$. Unfortunately, such a lift need not exist\footnote{As an example, consider the identity function $\phi:S^1\rightarrow S^1:e^{i\theta}\mapsto e^{i\theta}$ on the circle with $\ell=2\pi$. Then a candidate lift would be $\alpha(\theta)=\theta$, which is not a well-defined function $S^1\rightarrow\mathbb{R}$.} unless $\phi$ maps all non-contractible loops in $M$ to a loop in $S^1$ with zero winding number \cite{Hatcher2001}. In particular, one can lift all fields in $\phi\in C^\infty(M,S^1)$ to fields $\alpha\in C^\infty(M)$ if $M$ is simply connected. If it isn't, one can replace $M$ with its universal cover. Both extensions to the universal covers of $M$ and $S^1$ can introduce ambiguities. Let us study them separately. 

Let us assume for the moment that $M$ is simply connected and let $x_0\in M$. The lift of a field $\phi\in C^\infty(M,S^1)$ to $\alpha\in C^\infty(M)$ is uniquely defined once $\alpha(x)$ has be specified such that $\phi(x_0)=e^{2\pi i\alpha(x_0)/\ell}$. The ambiguity in the lift is then because one should identify $\alpha\sim\alpha+\ell$. Given that these two fields are completely equivalent in the physical theory, one can think of this transformation as a $\mathbb{Z}$ global gauge symmetry. A gauge fixing procedure must thus be implemented to fix the ambiguity. In order to do this, we can introduce a gauge fixing function $F:C^\infty(M)\rightarrow\mathbb{R}$ and implement the gauge fixing condition that $F(\alpha)\in [0,\ell)$. An obvious example suggested by the discussion at the beginning of the paragraph is $F(\alpha):=\alpha(x_0)$. However, unless $M$ has a distinguished point, it is not geometrically very natural. A more natural choice in these cases might be the average
\begin{equation}
    F(\alpha)=\frac{1}{\Vol M}\int\dd[D]{x}\sqrt{g}\alpha.
\end{equation}

Now, let us study the situation when $M$ is not simply connected. As long as it is connected and locally simply connected, it will admit a simply connected cover, its universal cover $p:\tilde{M}\rightarrow M$. This projection allows us to present $M$ as a quotient 
\begin{equation}
    M = \faktor{\tilde{M}}{\pi_1(M)}.
\end{equation}
We can then replace $M$ with a simply connected fundamental domain of this quotient, which can be done as long as one imposes the necessary boundary conditions on the boundary of this fundamental domain to ensure that the fields on the fundamental domain correspond to fields on $M$. To make the discussion as concrete as possible, let us consider specific examples of $M$ on which we can compute our path integrals.

\subsubsection{The \texorpdfstring{$D$}{D}-Torus}

Let us consider
\begin{equation}
    M=T^D=\underbrace{S^1\times\cdots\times S^1}_{D\text{ times}}.
\end{equation}
where the circles have circumferences $L=(L_1,\dots,L_D)$. Its fundamental cover is $\mathbb{R}^D$ and a fundamental domain of the quotient $\faktor{\mathbb{R}}{L\cdot\mathbb{Z}^D}$ can be taken to be any product of intervals of sizes $L_1,\dots,L_D$, say $[0, L]:=[0, L_1]\times\cdots\times[0, L_D]$. Then we can replace $C^\infty(S^1, S^1)$ by $C^\infty([0, L], S^1)$ as long as we further demand that our fields and all of its derivatives are periodic. These can be further lifted to the vector space $C^\infty([0, L])$. In this lift, the boundary conditions get lifted so that the derivatives remain periodic. The field, on the other hand, need not be periodic. Instead, it should satisfy $\phi(x+L_ie_i)=\phi(x)+n_i\ell$ for some fixed $n=(n_1,\dots,n_D)\in\mathbb{Z}^D$ known as the winding numbers of $\phi$. One obtains an identification of $C^\infty(T^D, S^1)$ with this space of fields as long as one fixes the gauge by the condition
\begin{equation}
    \frac{1}{\Vol M}\int_{[0,L]}\dd[D]{x}\phi(x)\in[0,\ell).
\end{equation}
In summary, we identify $C^\infty(T^D, S^1)$ with the space of fields $\mathcal{E}\subseteq C^\infty(N)$ defined by these boundary and gauge fixing conditions. In this way, we have transferred the non-linearity of the space of fields from the target $S^1$ to the boundary conditions.

In order to find a vector bundle structure for the space of fields, let us begin by noticing that the instantons\footnote{The instantons satisfy $\Delta\phi=0$. Accordingly, they correspond to null eigenfields of the Laplacian. These correspond to products of eigenfields of the Laplacians in each direction, corresponding to eigenvalues whose sum vanishes. However, the periodic boundary conditions prohibit any of these eigenvalues from being negative since that would lead to exponential solutions. Accordingly, all of the eigenvalues must vanish, meaning that they must be products of linear functions in each direction. Finally, for the winding boundary condition to be satisfied, there can only be one non-constant direction in each such product.} are of the form
\begin{equation}
    \phi_{n,a}=a + \frac{\ell n}{L}\cdot x-\sum_{i=1}^D\frac{\ell n_i}{2}\qc \frac{n}{L}:=(n_1/L_1,\dots,n_D/L_D),
\end{equation}
with an action
\begin{equation}
    S(\phi_{n,a})=\sum_{i=1}^D\frac{(\ell n_i)^2}{2L_i}.
\end{equation}
Here, $a$ is the average restricted by our gauge fixing condition,\footnote{It is an interesting exercise to instead take as a gauge fixing condition $\phi(0)$ and seeing how the discussion changes.} so that we can identify the space of instantons with $\mathcal{J}=[0,\ell)\times \mathbb{Z}^D$. We then define the vector bundle $\pi:\mathcal{E}\rightarrow [0,\ell)\times\mathbb{Z}^D$ by mapping every field to the instanton with the same average and winding number. The fibers are then of the form
\begin{equation}
    \phi_{n,a}+\mathcal{F},
\end{equation}
where $\mathcal{F}\subseteq \mathcal{E}$ are the functions with vanishing average and winding number. In particular, the vanishing of the winding number indicates that these are precisely the types of functions that can be thought of as functions on the torus $C^\infty(T^D)$ with vanishing average.

The space $C^\infty([0, L])$ can be endowed with our standard bosonic inner product \eqref{eq:inner_product_Klein}. This induces a measure on our base space given by the dimensionless combination
\begin{equation}
    \sum_{n\in\mathbb{Z}^D}\int\dd{a}\sqrt{\frac{\Vol M}{2\pi\mu}}.
\end{equation}
On the other hand, the beta function is obtained from the operator $Q$ on $\mathcal{F}$. This operator coincides with the operator $Q$ on $C^\infty(T^D)$ with its zero mode removed. Indeed, this is the mode corresponding to the constant function, which fixes the field average, which vanishes on $\mathcal{F}$. Therefore,
\begin{equation}
    \zeta_{\mu Q_{\mathcal{F}}}(s)=\tr((\mu Q_{\mathcal{F}})^{-s})=\zeta_{\mu Q_{C^\infty(T^D)}}(s)-(\mu m^2)^{-s},
\end{equation}
In the limit as $m\rightarrow 0$. This zeta function has already been explored in sections \ref{sec:scalar_closed} and \ref{app:scalar_torus}. In particular, for $D$ odd
\begin{equation}
    \beta = -1/2,
\end{equation}
so that $\beta+\dim\mathcal{J}/2=0$ and the path integral self-normalizes. 

The full path integral can also be computed using the results in \ref{app:scalar_torus}. For example, for $D=1$, we have
\begin{equation}
    S_\eff=\ln(2\sinh(mL/2))-\frac{1}{2}\ln(\mu m^2)+S(\phi_{n,a})\rightarrow \ln(L/\sqrt{\mu})+\frac{\ell^2 n^2}{2L}.
\end{equation}
Then, the path integral is given by
\begin{equation}\label{eq:partition_compact_boson}
    \int\mathcal{D}\phi\,e^{-S}=\sum_{n\in\mathbb{Z}}\int_0^{\ell}\dd{a}\sqrt{\frac{L}{2\pi\mu}}\frac{\sqrt{\mu}}{L}e^{-\ell^2n^2/2L}=\frac{\ell}{\sqrt{2\pi L}}\theta_3\qty(e^{-\ell^2/2L}),    
\end{equation}
where $\theta_3$ is the third theta function.

In $D=1$, we can also add a topological term to the action of the form
\begin{equation}
    \frac{i\theta}{\ell}\int\dd{\phi}, 
\end{equation}
for a dimensionless constant $\theta\sim\theta+2\pi$. While this term does not affect the equations of motion and thus leaves the instantons intact, it does provide an additional contribution to the effective action through the instanton action. 
\begin{equation}
    S_\eff=\ln(L/\sqrt{\mu})+\frac{\ell^2 n^2}{2L}+i\theta n.
\end{equation}
This leaves us with the path integral
\begin{equation}
    \int\mathcal{D}\phi\,e^{-S}=\sum_{n\in\mathbb{Z}}\int_0^{\ell}\dd{a}\sqrt{\frac{L}{2\pi\mu}}\frac{\sqrt{\mu}}{L}e^{-\ell^2n^2/L}=\frac{\ell}{\sqrt{2\pi L}}\vartheta\qty(\frac{\theta}{2\pi};\frac{i\ell^2}{2\pi L}),   
\end{equation}
where $\vartheta$ is the Jacobi theta function.

\subsubsection{Transition Amplitude}

Now, let us consider the simplest cylinder, namely, $M=[0, L]$. This space is already simply connected. Thus, we can lift the space of fields to $C^\infty([0, L])$. After these, the boundary conditions are fixed only up to $\ell$. We may, however, fix the gauge by fixing an initial value $\varphi_i$, say in $[0,\ell)$ and then demanding that $\phi(0)=\varphi_i$ and $\phi(L)=\varphi_f+n\ell$ for some $n\in\mathbb{Z}$. For each $n$, we thus have the case of a transitioning free particle, whose path integral is the $m\rightarrow 0$ limit of \eqref{eq:transition_harmonic_oscillator}. In particular, for each $n$, we have an instanton, i. e., the space of instantons is $\mathcal{J}=\mathbb{Z}$. The path integral is therefore
\begin{equation}
    \int\mathcal{D}\phi\,e^{-S}=\mu^{1/4}\sqrt{\frac{1}{2L}}\sum_{n\in\mathbb{Z}}e^{-\frac{(n\ell+\varphi_f-\varphi_i)^2}{2L}}.
\end{equation}
In particular, we can join the two ends of the interval to recover the path integral over the circle
\begin{equation}
    \int\frac{\dd{\varphi}}{\nu^{1/4}}\qty(\int\mathcal{D}\phi\,e^{-S})_{\varphi_i=\varphi_f=\varphi}=\qty(\frac{\mu}{\nu})^{1/4}\ell\sqrt{\frac{1}{2L}}\theta_3(e^{-\ell^2/2L}).
\end{equation}
We once again recover that
\begin{equation}
    \frac{\mu}{\nu}=\frac{1}{\pi^2},
\end{equation}
through comparison with the self-normalizing path integral \eqref{eq:partition_compact_boson}.

\subsection{Chiral Anomaly}

We will now consider a theory in an infinite volume Euclidean spacetime $\mathbb{R}^D$. Accordingly, our main focus is not to fix $\mu$. However, our discussion will help us show how understanding a theory's $\mu$ dependence can help us understand some of its features. In particular, this scheme is useful when understanding anomalies (see \cite{Ramond1990} for an example concerning the conformal anomaly of $\phi^4$ theory or \cite{Osgood1988} for a similar example in the case of the bosonic string). Let us consider a massless Dirac field in even dimension $D$ coupled to a background gauge field
\begin{equation}
    e^{-S_\eff(\mu, A)}:=\int\mathcal{D}\bar{\psi}\mathcal{D}\psi\, e^{-S(\bar{\psi},\psi, A)}\qc S(\bar{\psi},\psi,A):=\int\dd[D]{x}\bar{\psi}\slashed{D}\psi.
\end{equation}
In here $\psi$ and $\bar{\psi}$ are Grassmann valued, $D_\mu=\partial_\mu+A_\mu$ is the covariant derivative, and we are using the Feynman slash notation $\slashed{A}:=\gamma^\mu A_\mu$ for Hermitian gamma matrices satisfying the Clifford algebra.

This theory is free (for the background field is non-dynamical), so the path integral is Gaussian, and we can compute it with the techniques developed in this paper. The only modification needed to account for the Grassmann nature of our fields is that we will now aim to extend the finite-dimensional formula
\begin{equation}
    e^{-S_\eff}=\int\dd[N]{\bar{\theta}}\dd[N]{\theta}e^{-\ev{\bar{\theta},A\theta}}=\det A
\end{equation}
to the infinite-dimensional case. Here $\bar{\theta}$ is in the dual space to $\theta$, and $\ev{\cdot,\cdot}$ is the duality pairing between the two. That this duality pairing is not unique in our setting is again enforced by dimensional considerations. As we will see, though, it is also useful to consider pairings that are not simple rescalings of the canonical duality pairing.

In our case, we will take the duality pairing to be\footnote{The positioning of the scale anticipates that it will be useful to consider analogs in which the scale is not given by a number but by a more general operator on the space of spinors.}
\begin{equation}
    \ev{\bar{\psi},\psi}_\mu=\int\dd[D]{x}\bar{\psi}\frac{1}{\mu}\psi,
\end{equation}
for a length scale $\mu$, with $[\mu]=-1$. Then
\begin{equation}
    S(\bar{\psi},\psi, A)=\ev{\bar{\psi},\mu\slashed{D}\psi}_\mu\qand e^{-S_\eff(\mu, A)}=\det(\mu\slashed{D}).
\end{equation}
We would like to understand the behavior of this effective action under axial transformations
\begin{equation}\label{eq:chiral_symmetry}
    \psi\mapsto e^{i\alpha\Gamma}\psi \qand\bar{\psi}\mapsto\bar{\psi}e^{i\alpha\Gamma},
\end{equation}
where $\Gamma$ is the matrix whose eigenvalue is $1$ on right-handed spinors and $-1$ on left-handed spinors
\begin{equation}
    \Gamma = i^{D(D-1)/2}\gamma^1\cdots\gamma^D    
\end{equation}
This is, of course, a symmetry of the classical action since for infinitesimal $\alpha$
\begin{equation}
    \delta S=\int\dd[D]{x}\partial_\mu\alpha j^\mu\qc j^\mu:=\bar{\psi}\gamma^\mu\Gamma\psi.
\end{equation}
However, it will be anomalous in the quantum theory because it doesn't preserve the duality pairing we chose
\begin{equation}
    \delta\ev{\bar{\psi},\psi}_\mu=2i\int\dd[D]{x}\bar{\psi}\frac{\alpha\Gamma}{\mu}\psi.
\end{equation}
This, of course, should be contrasted with the vector symmetry $\psi\mapsto e^{i\alpha}\psi$ and $\bar{\psi}\mapsto \bar{\psi}e^{-i\alpha}$, which preserves both the classical action and the duality pairing and is, therefore, non-anomalous. Indeed, the duality pairing has the same form as a mass term in the action. While massive fermions retain the vector symmetry, such terms break its axial counterpart, even at the classical level. The fact that our approach has made manifest the dependence on the duality pairing of the theory provides a straightforward explanation of this anomaly.

The effect on the integral measure can thus be equivalently understood through a change in the duality pairing prescribed by
\begin{equation}
     \delta\mu =-2i\alpha\Gamma\mu,
\end{equation}
as opposed to a change in the field content. Note that even if we start with a scalar $\mu$, studying this symmetry forces us to consider more general pairings. We would like to understand how this affects the effective action. 

First, let us note that because the theory is of first order, the operator in the determinant is not a positive definite Laplacian. In order to use the methods of this paper, we can, however, formally replace
\begin{equation}
    \det(\mu\slashed{D})=\sqrt{\det(-\mu^2\slashed{D}^2)},
\end{equation}
up to a multiplicative phase. This phase is in itself complicated, and we will leave it outside of the scope of this work \cite{Witten2016, Asada2000}. In any case, after this replacement, we obtain the effective action defined through zeta function regularization
\begin{equation}
    S_\eff(\mu, A)=-\frac{1}{2}\tr\ln(-\mu^2\slashed{D}^2)=\frac{1}{2}\zeta_{-\mu^2\slashed{D}^2}'(0).
\end{equation}

The variation of the effective action due to the variation in the measure according to the axial anomaly is then given by the variation of the zeta function. 
\begin{equation}
\begin{aligned}
    \delta\zeta_{-\mu^2\slashed{D}^2}(s)=4is\tr(\alpha\Gamma(-\mu^2\slashed{D}^2)^{-s}).
\end{aligned}
\end{equation}
After analytically extending the trace, we obtain then
\begin{equation}
    \delta S_\eff(\mu, A)=2i\tr(\alpha\Gamma(-\mu^2\slashed{D}^2)^{-s})|_{s=0}.
\end{equation}
We are thus left with the same problem we've analyzed before of analytically continuing a zeta function at $s=0$. As before, we can perform the required analytic continuation by relating the resulting zeta function to a heat kernel
\begin{equation}
    \tr(\alpha\Gamma(-\mu^2\slashed{D}^2)^{-s})=\frac{1}{\Gamma(s)}\int_0^\infty\dd{t}t^{s-1}\tr(\alpha\Gamma e^{t\mu^2\slashed{D}^2}).
\end{equation}
Then, the value of this at $s=0$ is given by the constant term of the $t\rightarrow 0^+$ asymptotic expansion of the trace on the right-hand side using the arguments from section \ref{sec:scale_D_2}.

At this stage, we have arrived at the computation of the Jacobian of the path integral measure found in the literature \cite{Schwartz2014} using the techniques developed in this paper. It is important to note that our method has further elucidated that the full computation of the heat kernel trace is not what is important but rather only the constant term of its asymptotic expansion. This explains why the result obtained in the literature at one loop is, in fact, exact. For completeness, we will now proceed to complete the computation. We will mainly follow the arguments presented in \cite{Tic}.

In order to understand the asymptotic expansion, we will first extract the $1/t^{D/2}$ dependence that we expect from the asymptotics of the standard heat kernel. In order to do this, we will expand the trace in plane waves
\begin{equation}
    \tr(\alpha\Gamma e^{t\mu^2\slashed{D}^2})=\int\frac{\dd[D]{k}}{(2\pi)^D}\ev{\alpha\tr(\Gamma e^{t\mu^2\slashed{D}^2})}{k},
\end{equation}
where the trace on the right-hand side is only over the spinorial indices. Since our operators are written in position space, we can further proceed by performing a change of basis into position space
\begin{equation}
    \tr(\alpha\Gamma e^{t\mu^2\slashed{D}^2})=\int\dd[D]x\alpha\int\frac{\dd[D]{k}}{(2\pi)^D} \tr(\Gamma e^{-ikx} e^{t\mu^2\slashed{D}^2}e^{ikx}).
\end{equation}
The operator in the trace can be further simplified using the relationship between the exponentiated and un-exponentiated adjoint actions
\begin{equation}
    e^{-ikx} e^{t\mu^2\slashed{D}^2}e^{ikx}=e^{t\mu^2e^{-ikx}\slashed{D}^2e^{ikx}}.
\end{equation}
One can then compute the spinor Laplacian past the plane wave to
\begin{equation}
    e^{-ikx}\slashed{D}^2e^{ikx}=(e^{-ikx}\slashed{D}e^{ikx})^2=(i\slashed{k}+\slashed{D})^2=-k^2+2ik^\mu D_\mu+\slashed{D}^2.
\end{equation}
Putting this together, we can rescale the $k$ variable to extract the desired divergence
\begin{equation}
    \tr(\alpha\Gamma e^{t\mu^2\slashed{D}^2})=\frac{1}{(\mu t)^{D/2}}\int\dd[D]{x}\alpha\int\frac{\dd[D]{k}}{(2\pi)^D}e^{-k^2}\tr(\Gamma e^{2ik^\mu\sqrt{t}\mu D_\mu+t\mu^2\slashed{D}^2}).
\end{equation}

In order to find the constant term, we then want to expand the exponential to order $\order{t^{D/2}}$. This means that the expansion must have $D$ derivatives $D_\mu$. On the other hand, all of these derivatives must be of the slashed kind $\slashed{D}$, for the only non-vanishing trace of a product of $\Gamma$ and the gamma matrices is the one in which all the gamma matrices appear  
\begin{equation}
    \tr(\Gamma\gamma^{\mu_1}\cdots\gamma^{\mu_D})=D(-i)^{D(D-1)/2}\epsilon^{\mu_1\dots\mu_D}.
\end{equation}
With these considerations in mind, the only contribution is
\begin{equation}
    \tr(\Gamma e^{2ik^\mu\sqrt{t}\mu D_\mu+t\mu^2\slashed{D}^2})\sim\frac{(t\mu^2)^{D/2}}{(D/2)!}\tr(\Gamma\slashed{D}^D).
\end{equation}
All of the $k$-dependence is now under control, and the corresponding integral can be computed. Furthermore, the remaining trace can be expressed in terms of the curvature of the gauge field $[D_\mu, D_\nu]=F_{\mu\nu}$, due to the antisymmetry of the Levi-Civita tensor
\begin{equation}
\begin{aligned}
    \tr(\Gamma\slashed{D}^D)&=D(-i)^{D(D-1)/2}\epsilon^{\mu_1\dots\mu_D}D_{\mu_1}\cdots D_{\mu_D}\\
    &=\frac{D(-i)^{D(D-1)/2}}{2^{D/2}}\epsilon^{\mu_1\dots\mu_D}F_{\mu_1\mu_2}\cdots F_{\mu_{D-1}\mu_D}\\
    &=D(-i)^{D(D-1)/2}\frac{F^{\wedge D/2}}{\dd[D]{x}},
\end{aligned}
\end{equation}
written in terms of the curvature $2$-form $F=\frac{1}{2}F_{\mu\nu}\dd{x^\mu}\wedge \dd{x^\nu}$. We conclude that the constant term is
\begin{equation}
    \tr(\alpha\Gamma e^{t\mu^2\slashed{D}^2})\sim\frac{D(-i)^{D(D-1)/2}}{(4\pi)^{D/2}(D/2)!}\int\alpha F^{\wedge D/2},
\end{equation}
i. e., the variation of the effective action due to the change of the path integral measure is
\begin{equation}
    \delta S_\eff = \frac{-2D(-i)^{D(D-1)/2+1}}{(4\pi)^{D/2}(D/2)!}\int\alpha F^{\wedge D/2}.
\end{equation}

This result is usually expressed as an anomaly in the conservation equation for the current $j^\mu$. In order to do this, one should note that \eqref{eq:chiral_symmetry} amounts to a transformation in the dummy variables of integration in the path integral. Accordingly
\begin{equation}
    0=\delta\qty(\int\mathcal{D}\bar{\psi}\mathcal{D}\psi\, e^{-S})=\int\delta\qty(\mathcal{D}\bar{\psi}\mathcal{D}\psi)\, e^{-S}+\int\mathcal{D}\bar{\psi}\mathcal{D}\psi\, \delta e^{-S}.
\end{equation}
The first summand corresponds precisely to the variation of the effective action due to the change in the path integral measure, i. e., the change in the inner product used to define it
\begin{equation}
    \delta\qty( e^{-S_\eff})=-e^{-S_\eff}\delta S_\eff.
\end{equation}
The second summand instead corresponds to the variation of the action under \eqref{eq:chiral_symmetry}
\begin{equation}
    \int\mathcal{D}\bar{\psi}\mathcal{D}\psi\, \delta e^{-S}=-\int\mathcal{D}\bar{\psi}\mathcal{D}\psi\, e^{-S}\int\dd[D]{x}\partial_\mu\alpha j^\mu.
\end{equation}
If this is to be true for arbitrary $\alpha$ of compact support, integration by parts then shows that we must have
\begin{equation}
    \dd{j} = \frac{-2D(-i)^{D(D-1)/2+1}}{(4\pi)^{D/2}(D/2)!} F^{\wedge D/2},
\end{equation}
for the associated $(D-1)$-form $j$. 

\section{Conclusions}

In this paper, we argued that the inner product structure on the space of field configurations, a key piece of data required for the quantization of a classical field theory, is the source of the ambiguity of the normalization of the path integral. We further found that this inner product structure requires the introduction of a scale $\mu$. The dependence on this scale can be obtained through a beta function $\beta$, which can, in turn, be computed using zeta function regularization techniques in conjunction with heat-kernel methods. We showed that the behavior of the zeta function is heavily constrained by the consistency conditions obtained by cutting and pasting different path integrals. We argued that a complete understanding of such procedures should fully determine the scale. 

In $D=1$, we were able to give explicit examples of this, being able to compute the beta function in a simple algebraic manner through an ``index-like'' expression \eqref{eq:index}, and obtain an explicit expression for the scale $\mu=\nu/\pi^2$. We further exemplified the methods by computing beta functions and full path integrals in different geometries for the free and compact scalar fields. Furthermore, we showed that the methods can be extended to the study of anomalies, making an explicit computation of the chiral anomaly. In particular, the methods give a simple proof that the anomaly is one loop exact.

Several future research directions stem from this work. The major limitation of this paper was the difficulty of computing boundary path integrals for $D> 1$. Due to this, we could only provide heuristic arguments for determining $\mu$ in these cases. This is, of course, an active area of research \cite{Freed1999}, and we believe our work will provide important hints towards its resolution. On a related note, even though we highlighted the resemblance of our $\mu$ with the scale appearing in dimensional regularization, in our work, we refrain from analyzing the role that $\mu$ will play in renormalization group flow. In particular, effective field theory suggests that $\mu$ will be a function of the energy scale of the theory, but we did not need an effective field theory interpretation of the scale to get our arguments off the ground.

Another future direction is to connect this work with another one of the disanalogies between QFT and SFT: the causal structure. Indeed, throughout this paper, we worked with ``Euclidean'' QFT, using the principles of QFT, but at the same time, we worked in an Euclidean space where the action was real. In real relativistic models, the spacetime is Lorentzian, leading to purely imaginary actions. More generally, state construction and measurements of observables, say, through the Schwinger-Keldysh formalism, requires consideration of complex metrics \cite{Witten2021}. The study of such path integrals requires a proper deformation of the integration contour into a complexification of the fields, say, by using Lefschetz thimbles \cite{wittenNewLookPath2010, Tanizaki2014}. It would be interesting to apply the methods studied in this paper to these path integrals.

\acknowledgments

We want to thank Andrés F. Reyes-Lega for helpful feedback in framing this paper and Bruno de S. L. Torres's suggestions that led to section \ref{sec:scale_D_2}. We would also like to thank Petr Hořava for useful comments on the manuscript, James D. Wells and Aleksandr Pinzul for guidance on presenting our work, and Chris Li for spotting a typo in the manuscript.

\bibliographystyle{unsrt}
\bibliography{references}

\end{document}